\documentclass{aastex61}
\usepackage{amsmath,amstext}
\usepackage{natbib}
\usepackage{bm}
\usepackage{multirow}
\usepackage{color}
\newcommand{\mitsuishi}{\textcolor{red}}

\received{\today}
\submitjournal{ApJ}

\shorttitle{Comparison of Solar and Stellar Coronae}
\shortauthors{Takasao et al.}

\begin{document}

\title{Investigation of coronal properties of X-ray bright G-dwarf stars based on the solar surface magnetic field -- corona relation}

\correspondingauthor{Shinsuke TAKASAO}
\email{takasao@astro-osaka.jp}

\author[0000-0003-3882-3945]{Shinsuke TAKASAO}
\affiliation{Department of Earth and Space Science, Osaka University, Toyonaka, Osaka 560-0043, Japan}

\author{Ikuyuki Mitsuishi}

\author{Takuma Shimura}

\author{Atsushi Yoshida}
\affiliation{Graduate School of Science, Division of Particle and Astrophysical Science, Nagoya University, Furo-cho, Chikusa-ku, Nagoya, Aichi 464-8602, Japan}

\author[0000-0002-1932-3358]{Masanobu Kunitomo}
\affiliation{Department of Physics, School of Medicine, Kurume University, 67 Asahi-machi, Kurume, Fukuoka 830-0011, Japan}

\author{Yuki A. Tanaka}
\affiliation{Astronomical Institute, Tohoku University, 6-3 Aramaki, Aoba-ku, Sendai, Miyagi, 980-8578, Japan}

\author{Daisuke Ishihara}
\affiliation{Institute of Space and Astronautical Science (ISAS), Japan Aerospace Exploration Agency (JAXA), 3-1-1 Yoshinodai, Chuo-ku, Sagamihara, Kanagawa 252-5210, Japan}



\begin{abstract}
We investigated the coronal properties of G-dwarf stars including the Sun over a wide range of X-ray luminosity $L_{\rm X}$ ($3\times 10^{26}$ to $2\times 10^{30}~{\rm erg~s^{-1}}$). We analyzed the archival data of ten X-ray bright ($L_{\it X}>10^{28}~{\rm erg~s^{-1}}$) G-dwarf stars to derive their emission measure (EM) and the coronal temperature ($T$) during the periods when no prominent stellar flares were observed. We attempted to explain the relation on the basis of our understanding of the present Sun: a steady corona model based on the so-called RTV scaling laws and the observed power-law distribution function of surface magnetic features. We derived a theoretical scaling law of the EM--$T$ relation for a star with multiple active regions, and applied it to the observations combined with data in literature. We found that with the solar parameters, our scaling law seems to be consistent with the data of slowly-rotating stars. However, more X-ray bright stars are located well above the scaling law based on the solar parameter. The scaling law may explain the observations if those stars show a power-law distribution function of active regions with the same power-law index but a 10-100 times larger coefficient. This suggests that X-ray bright stars show more active regions for a given size than the Sun. Since our samples include rapidly-rotating stars, we infer that the offset of the X-ray bright stars from the present-Sun-based scaling law is due to the enhancement of the surface magnetic field generation by their rapid rotation.
\end{abstract}


\section{Introduction}
Stellar coronae are atmospheres of hot plasma at temperatures exceeding one million K. Dynamical processes in stellar coronae are known to be important not only for the stellar evolution but also the evolution of surrounding planets. Stellar winds blow out from coronae as a result of the energy injection from the stellar surface \citep{sheeley1997,suzuki2005,cranmer2007}, and regulate the evolution history of the stellar spin and mass \citep{weber1967,lamers1999}. Stellar winds can also dynamically influence the atmospheres of planets including the Earth \citep{cohen2015,shiota2016}. Stellar flares, which are explosive magnetic energy releases, are another important source of the disturbance \citep{Tsurutani2005,Airapetian2016,airapetian2019}. High-energy radiations from the stellar coronae such as X-rays and EUV photons can also affect the planet atmospheres \citep{linsky2019}. Such high-energy radiations can cause planetary atmospheres to expand and eventually evaporate \citep{lammer2003,Lalitha2018}. Because of those different impacts, it is crucial to reveal what determines the properties of the stellar coronae. What we clearly know is that the stellar coronal heating is tightly coupled with magnetic energy release.




The signature of magnetic energy release is most prominently seen in coronal emissions like X-ray. Although optical observations can provide the information about photospheric magnetic fields such as starspots, the interpretation is sensitive to the inclination angle to the line-of-sight \citep{ynotsu2013a}. On the other hand, coronal X-rays are emitted from optically-thin, large volumes above active regions, and therefore the inclination effect is less significant. The energy of solar and stellar flares can be well constrained by X-ray observations.
For these reasons, X-ray observations have been extensively performed to understand the magnetic activities in the stellar atmospheres \citep{Gudel2004,Testa2015,tsuboi2016}. Previous studies suggest that the X-ray luminosity of the non-flaring (or quasi-steady) coronae can be as high as $10^{-3}L_{\rm bol}\approx 4\times 10^{30}$~erg~s$^{-1}$ for G-dwarf stars \citep{Gudel1997, Gudel2004}, where $L_{\rm bol}$ is the stellar bolometric luminosity. This value is much larger than the solar X-ray luminosity; $5\times 10^{27}$~erg~s$^{-1}$ at maximum \citep{Peres2000}. A fraction of those X-ray bright G-dwarf stars will correspond to solar-type superflare stars with huge starspots \citep{maehara2012,shibata2013}, although there are few simultaneous observations in optical and X-ray \citep{YNotsu2017}.

The central questions of this paper are as follows: {\it Can we derive a scaling law that describes the solar coronal properties? Does the Sun-based scaling law describe the coronal properties of other G-dwarf stars?}
To answer these questions, we need a set of single/wide binary stars with a wide range of coronal parameters such as the X-ray luminosity and the coronal temperature, to see the trend. Selecting single stars and wide binaries only will also be important for this aim, because in close binaries, the stellar dynamo will be significantly affected by the tidal effect. Our understanding of single X-ray bright G-dwarf stars is quite limited due to the small number of samples. \citet{Gudel1997} investigated the coronal property of solar-type G stars at different ages, and found that the coronal temperature increases as $T\propto L_{\rm X}^{1/4}$, where $T$ and $L_{\rm X}$ are the coronal temperature and the X-ray luminosity, respectively \citep[see also][]{telleschi2005,johnstone2015}. However, there is only one single star in the range of $L_{\rm X}>10^{29}$~erg~s$^{-1}$ in their plots. We therefore lack the observational information about single/wide binary, X-ray luminous G-dwarf stars.

The X-ray properties of stars are determined by the contribution of multiple active regions. One will notice that the coronal structures in X-ray is highly nonuniform and X-ray emissions are localized in active regions. Therefore, for understanding the link between the coronal emissions and the surface magnetic fields, it is crucial to consider the spatial distribution of the surface magnetic fields. The distribution for the Sun has been measured in detail \citep{parnell2009}. The property of the surface magnetic fields will depend on the stellar rotation period, as we empirically know that X-ray luminosity is almost linearly proportional to the total unsigned magnetic flux \citep{Pevtsov2003} and rapidly rotating stars are more luminous in X-ray than slowly rotating stars \citep{2003A&A...397..147P}. However, a quantitative discussion about the distribution of the surface magnetic fields is missing.

As a first step to answer the above question, we analyzed the archival data of G-dwarf stars, and searched for X-ray bright stars with $L_{\rm X}\gtrsim 10^{28}~{\rm erg~s^{-1}}$ to increase the number of samples of X-ray bright stars. In this paper, we present the observed coronal EM--$T$ relation of X-ray bright G-dwarf stars, and compare the solar and stellar coronae. For a quantitative comparison, we derive a scaling law of EM--$T$ relation for a star with multiple active regions, based on the current understanding of the present Sun.
The remainder of the paper is structured as follows.
Section~\ref{sec:theory} reviews previous theories of the solar and stellar coronae, and presents the new scaling law. In the derivation, we utilize the distribution function of photospheric magnetic features observed in the present Sun. Our scaling law enables us to compare the Sun and other G-dwarf stars. Section~\ref{sec:observations} explains the procedure of our sample selection, data reduction, and data analysis. We will describe how we derive the X-ray luminosity, the coronal temperature, and the emission measure for our targets from the data taken by {\it XMM-Newton} X-ray observatory. The rotation periods of our targets are shown in Section~\ref{sec:rotation}, although the period are estimated for handful stars only. In Section~\ref{sec:theory-obs-comp}, we apply our theoretical scaling law to the observational dataset that combines existing samples and our new samples, and compare the solar and stellar coronae. Section~\ref{sec:summary-discussion} summarizes and discusses our results.

\section{Derivation of Integrated Coronal Quantities based on Solar Corona Theory and Solar Observations}\label{sec:theory}
\subsection{Scaling Relations for Single Active Regions}
Stellar coronae are heated by the energy injection from the photosphere. The coronae are simultaneously subject to the cooling by the energy transport via the heat conduction. The heat transported to the lower coronae is finally taken away by the radiative cooling. Therefore, three energy fluxes (a heating flux, the thermal conduction flux, and the radiative cooling flux) are involved. In a steady state, the magnitude of each energy flux should be the same. The energy balance determines the thermal properties of the stellar coronae. Under such a condition, we can derive some useful and general scaling relations. We briefly review the discussions of \citet{rosner1978}, the first paper to derive scaling relations based on the energy balance, and \citet{Shibata2002}. We then derive some new relations for further discussions.

We assume that a stellar corona is in a steady state in which the heating flux $F_{\rm h}$, the thermal conduction flux $F_{\rm c}$, and the radiative cooling flux $F_{\rm r}$ are all comparable:
\begin{align}
F_{\rm h} \approx F_{\rm c} \approx F_{\rm r} ~{\rm erg~cm^{-2}~s^{-1}}.
\end{align}
Since the targets analyzed in this study do not show strong X-ray variability (see Section~\ref{sec:analysis}), we consider that the assumption of the quasi-steady coronae is valid. The explicit forms are $F_{\rm c}=|\kappa \partial T/\partial s|$ and $F_{\rm r} = n^2 \Lambda(T)l$. $\kappa$ is the Spitzer conduction coefficient $\kappa = \kappa_0 T^{5/2}$, where $\kappa_0$ is a constant ($\sim 10^{-6}$ in cgs units). $s$ is a spatial coordinate along a magnetic loop. $n$ is the number density. $\Lambda(T)$ is the radiative loss function in $\rm erg~cm^3~s^{-1}$. $l$ is the typical length of coronal magnetic loops of interest.

\begin{figure*}
\epsscale{0.5}
\plotone{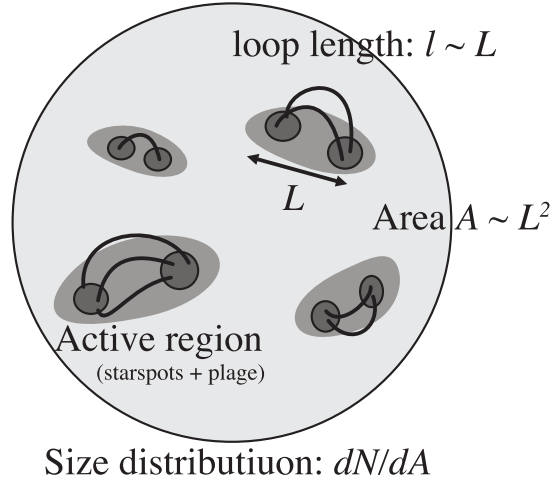}
\caption{Schematic diagram of the stellar surface with multiple active regions. Solid lines denote magnetic field lines. Dark gray regions denote starspots, while the surrounding light gray areas indicate plages.}\label{fig:model}
\end{figure*}

The thermal conduction flux $F_{\rm c}$ can be approximated as $F_{\rm c} \approx \kappa T/l \approx \kappa_0 T^{7/2}/l$. We therefore obtain
\begin{align}
F_{\rm h}&\approx \kappa_0 T^{7/2}l^{-1}~{\rm erg~cm^{-2}~s^{-1}} \label{eq:Fh-Tl}\\
\left(\frac{l}{10^{10}~{\rm cm}}\right)&\approx 10^{-2}\left(\frac{T}{10^6~{\rm K}}\right)^{7/2} \left(\frac{F_{\rm h}}{10^7~{\rm erg~cm^{-2}~s^{-1}}}\right)^{-1}\label{eq:l-T}.
\end{align}
According to \citet{Withbroe1977}, the typical value of $F_{\rm h}$ for solar active regions is $10^{6-7}~{\rm erg~cm^{-2}~s^{-1}}$, and we take $10^{7}~{\rm erg~cm^{-2}~s^{-1}}$ in the following.
It is reasonable to consider that the loop size is comparable to the size of the active regions (see a schematic diagram of active regions in Figure~\ref{fig:model}. We will mention this point later again). If we assume that the heating flux $F_{\rm h}$ does not depend on $l$ nor $T$, Equation~\ref{eq:l-T} describes the relation between the size of active regions and the coronal temperature for a given heating flux. Indeed, the heating flux will not explicitly depend on these quantities both in the Alfv\'en wave heating scenario and the nanoflare heating scenario, which have been commonly considered for the solar coronal heating problem. 
For instance, see the discussion by \citet{Parker1988} about the nanoflare heating scenario. We note that a dependency of the heating flux on the magnetic field strength may be important \citep{2020arXiv200613978Z}.


We can also derive a scaling relation for the density. The balance between the conduction flux and the radiative cooling leads to the following relation:
\begin{align}
n\approx 10^{6.5} T^2 l^{-1}~{\rm cm^{-3}}\label{eq:RTV}
\end{align}
for $T<10^7~$K, where we used an approximated form of the radiative loss function, $\Lambda(T)\approx 3\times 10^{-23}(T/10^7~{\rm K})^{-1/2}$ for $T<10^7~$K (as we will see later, our data show that observed coronal temperatures are at most $10^7$~K). We neglected the dependence of $\Lambda(T)$ on metallicity for simplicity \citep[see][for more detail]{sutherland1993}. This scaling is essentially the same as the scaling law of Rosner-Tucker-Vaiana (RTV scaling) \citep{rosner1978}. It has been confirmed that the dependence of physical quantities of active regions is consistent with the RTV scaling and Equation~\ref{eq:l-T} \citep[e.g.][]{yashiro2001}. The RTV scaling has been widely applied to different astrophysical systems \citep[e.g.][]{2017ApJ...847...46T}.

We can utilize Equations~\ref{eq:l-T} and \ref{eq:RTV} to derive physical quantities for an active region with the spatial scale of $L$ \citep{Shibata2002}. Let us write the emission measure $\rm EM$ as 
\begin{align}
{\rm EM} = f n^2 L^3~{\rm cm^{-3}},\label{eq:EM}
\end{align}
where $L$ is the size of the active region, and $f$ is the filling factor of the corona with the number density $n$ and the temperature $T$ in the volume of $L^3$. See the schematic diagram of the stellar surface with multiple active regions in Figure~\ref{fig:model}, where the correspondence of $l$ and $L$ is shown. Active regions consist of starspots and plages, where plages are bright regions in the chromosphere and associated with a stronger magnetic field than the quiet regions. In the solar case, the ratio of plage to sunspot is approximately 10 \citep{1997ApJ...482..541C}. The assumption that the volume can be approximated as $L^3$ is roughly supported by \citet{aschwanden2008a}. By assuming that active region coronae are a collection of multiple magnetic loops with the size of $L$ and that $l\approx L$ \citep[e.g.][]{aschwanden2008b}, we can obtain the following expression of EM for a single active region from Equations~\ref{eq:l-T}, \ref{eq:RTV}, and \ref{eq:EM}:
\begin{align}
{\rm EM_{\rm sin}} &\approx 10^{44}\left(\frac{f}{0.1}\right) \left( \frac{T}{10^6~{\rm K}}\right)^{15/2} \left(\frac{F_{\rm h}}{10^7~{\rm erg~cm^{-2}~s^{-1}}}\right)^{-1}~{\rm cm^{-3}}\label{eq:EM-singleAR-TF}\\
&\approx 10^{338/7} \left(\frac{f}{0.1}\right) \left(\frac{F_{\rm h}}{10^7~{\rm erg~cm^{-2}~s^{-1}}}\right)^{8/7}\left(\frac{L}{10^{10}~{\rm cm}}\right)^{15/7}~{\rm cm^{-3}}\label{eq:EM-singleAR}\\
&\approx 10^{338/7} \left(\frac{f}{0.1}\right) \left(\frac{F_{\rm h}}{10^7~{\rm erg~cm^{-2}~s^{-1}}}\right)^{8/7}\left(\frac{A}{10^{20}~{\rm cm^2}}\right)^{15/14}~{\rm cm^{-3}}\label{eq:ShibataYokoyama}
\end{align}
for $T<10^7~\rm K$, where we assume that the area of the active region $A$ is given as $A\approx L^2$. $f=0.1$ is a reasonable value for the heating flux of active region coronae to become consistent with observations \citep{Shibata2002}.

\subsection{Scaling Relation for a Star with Multiple Active Regions}
Equation~\ref{eq:ShibataYokoyama} is for a single active region. However, stars generally have multiple active regions that obey a size distribution function (see Figure~\ref{fig:model}). Therefore, what we observe are the integrated quantities over such multiple active regions. Using a size distribution function $dN/dA(A)$, the number of active regions with the area $A$ per stellar surface area, we can write the total emission measure as follows:
\begin{align}
{\rm EM}_{\rm tot} = \int^{A_{\rm max}}_{A_{\rm min}} {\rm EM_{sin}}(A)\frac{dN}{dA} dA, \label{eq:EM-beforeInteg}
\end{align}
where $A_{\rm min}$ and $A_{\rm max}$ are the areas of the active regions with the minimum and maximum sizes, respectively.

The size distribution function $dN/dA$ for the Sun is observationally derived.
\citet{parnell2009} showed that the distribution of magnetic features (including not only active regions but also small magnetic patches) obey a single power law \citep[see also][]{2012ApJ...752..149I}. Their result is 
\begin{align}
\frac{d^2N}{d\phi dA}(\phi) = N_{\rm f}\phi^{-\alpha} ~{\rm Mx^{-1}~cm^{-2}}
\end{align}
where $d^2N/d\phi dA(\phi)$ is the distribution function of the magnetic features with the unsigned magnetic flux $\phi$ per area, $\alpha=1.85\pm 0.14$, and $N_{\rm f} = 3\times 10^{-4}$ in cgs units. The distribution function per total solar surface area can be obtained by multiplying the total surface area, $A_\odot = 4\pi R_\odot^2 \approx 6.2\times 10^{22}~{\rm cm^2}$:
\begin{align}
\frac{dN}{d\phi} = A_\odot N_{\rm f}\phi^{-\alpha} ~{\rm Mx^{-1}}.
\end{align}
With the assumption that $\phi \approx \bar{B}A$, where $\bar{B}$ is the typical magnetic field strength of the active regions, we obtain
\begin{align}
\frac{dN}{dA} &= A_\odot N_{\rm f}\bar{B}^{-\alpha+1}A^{-\alpha} ~{\rm cm^{-2}}\\
&\equiv \gamma_0  A^{-\alpha}~{\rm cm^{-2}},\label{eq:solar-size-dist}
\end{align}
where $\gamma_0 = A_\odot N_{\rm f}\bar{B}^{-\alpha+1}$.

As the distribution function for other stars, we consider the following function:
\begin{align}
    \frac{dN}{dA}=N_0\gamma_0 A^{-\alpha}~{\rm cm^{-2}}\label{eq:stellar-size-dist}
\end{align}
Here we hypothesize that the size distribution functions of other G-dwarf stars have the same power-law index as the solar one, but the difference appears in the coefficient. $N_0=1$ corresponds to the Sun, and $N_0>1$ corresponds to a star that shows more active regions with a given size than the Sun.

Equation~\ref{eq:EM-beforeInteg} suggests that $\rm EM_{tot}$ depends on both the minimum and maximum sizes of active regions. However, we will find that $\rm EM_{\rm tot}$ almost depends only on the maximum size, because $\alpha < 2$. The result of the integration based on the above size distribution (Equation~\ref{eq:stellar-size-dist}) is as follows:
\begin{align}
{\rm EM}_{\rm tot} &\approx \frac{14N_0\gamma_0}{29-14\alpha} 10^{60-16\alpha} \left( \frac{f}{0.1}\right)
\left( \frac{F_{\rm h}}{10^7~{\rm erg~cm^{-2}~s^{-1}}}\right)^{-3+2\alpha} \left( \frac{T_{\rm max}}{10^{6}~{\rm K}}\right)^{(29-14\alpha)/2}~{\rm cm^{-3}}\label{eq:EM-afterInteg}
\end{align}
where we use Equation~\ref{eq:l-T} to transform the size $L_{\rm max}$ to the temperature $T_{\rm max}$ (therefore, the temperature here denotes the temperature of the largest active region). 

We note that the coronal temperature derived from observations $T_{\rm obs}$ should be interpreted as
\begin{align}
T_{\rm obs} = \frac{\int_{A_{\rm min}}^{A_{\rm max}} {\rm EM}_{\rm sin}(A)\frac{dN}{dA} T(A)}{\int_{A_{\rm min}}^{A_{\rm max}} {\rm EM}_{\rm sin}(A)\frac{dN}{dA}}.
\end{align}
Using the relations~\ref{eq:l-T} and $A\sim l^2$ and the fact that $\alpha < 2$, we obtain
\begin{align}
    T_{\rm obs} = \frac{29-14\alpha}{31-14\alpha}T_{\rm max} \approx 0.61 T_{\rm max}\label{eq:Tobs}
\end{align}
for $\alpha=1.85$. We use relations~\ref{eq:EM-afterInteg} and \ref{eq:Tobs} to compare our scaling relation with observations. In the following, we fix $\alpha$ to be the present solar value, 1.85.

\section{Observations}\label{sec:observations}
\subsection{Sample selection}\label{sec:sample}
We search for X-ray bright G-dwarf stars by cross referencing the Tycho-2 spectral catalog \citep[e.g.][]{2003AJ....125..359W} to the third {\it XMM-Newton} serendipitous source catalogs \citep[e.g.][]{2016A&A...590A...1R}. These catalogs include $>$300,000 and $>$700,000 sources, respectively. To avoid matching failure, the cross-referencing was performed only for stars located outside the Galactic plane ($|$b$|$ $>$ 10 degrees). The search radius was set to 10~arcseconds.
As the first set of samples, we picked up stars with a luminosity class of V and a spectral type of G, and with an X-ray luminosity of \mitsuishi{$\gtrsim$}10$^{29}$~erg~s$^{-1}$ based on the flux in the catalog. As a result, we obtained only two G-dwarf stars (X-ray bright stars are rare). The basic information is summarized in Table \ref{tab:obssummary1}.

The number of the first set of samples is too small to discuss relations among stellar coronal parameters. To increase the sample number, we prepare another set of samples in the following manner. The search rules for new samples are the same except for the constraint on the X-ray luminosity. We looked for stars in our catalog whose photon statistics are high enough for the X-ray spectroscopic analyses. As a result, we obtained eight G-dwarf stars. The basic information of this second set of samples is summarized in Table~\ref{tab:obssummary2}.

We removed binary systems and possible non G-dwarf stars in previous observations by utilizing the SIMBAD astronomical database and GAIA DR2 database \citep{2019A&A...623A..72K}. As an exception, we include a binary star, named as Target~3 in our sample, because the separation distance is so large \citep[65 arcseconds, or 1700--2600 AU,][]{2007AJ....133..439C} that the stellar dynamo will not be affected by the companion. Also, our spectral analysis is not affected by the companion star due to the large separation distance (see Section~\ref{sec:analysis} for the detail about the spectral analysis).

\begin{table*}[htb]
 \caption{Basic information and logs of observations for two X-ray bright G-dwarf stars.}
  \begin{center}
\begin{tabular}{lllllllll}
\hline \noalign{\vskip2pt} 
Target Name & Target ID$^a$ & R.A.$_{{\rm J2000}}$     &   Dec.$_{{\rm J2000}}$   & SpT$^b$   & T$_{\rm{eff}}$$^c$ & dist.$^c$ &  Obs. ID$^d$     &  Exp.$^e$  \\
           & [deg.]     & [deg.]                    &  & [K]   &  [pc]  &  &  [ks]     \\ \hline
\multirow{3}*{Target 1} & \multirow{3}*{\shortstack{\\3XMMJ041422.7-381900 \\(TYC 7578-414-1)}} & \multirow{3}*{63.595} & \multirow{3}*{-38.317}  & \multirow{3}*{G3V} & \multirow{3}*{5860} & \multirow{3}*{80}
& 0677181001    &   10   
\\ \cline{8-9}
&   &    &      &   &   &  & 0720251501    &   10 
\\ \cline{8-9}
&   &    &      &   &   &   & 0720253501    &   24   
\\ \hline
Target 2 & \shortstack{\\3XMMJ143111.5-001415 \\(TYC 4984-18-1)} & 217.798 & -0.238   & G1V & 6020 & 98 & 0501540201                          &     16     \\ \hline
\end{tabular}
\label{tab:obssummary1}
\begin{flushleft} 
\footnotesize{
\hspace{0.5cm}$^a$ Target name in the X-ray and optical catalogs tabulated in \citet{2016A&A...590A...1R} and \citet{2003AJ....125..359W}. \\
\hspace{0.5cm}$^b$ Spectral type based on the SIMBAD astronomical database. \\
\hspace{0.5cm}$^c$ Effective temperature and distance are shown in \citet{2016A&A...595A...1G,2018AJ....156...58B}, respectively.\\
\hspace{0.5cm}$^d$ The XMM-Newton observation identification. \\
\hspace{0.5cm}$^e$ Net exposure time of the PN or MOS detector after removal of periods of high background flaring.}
\end{flushleft}
  \end{center}
\end{table*}

\begin{table*}[htb]
 \caption{Basic information and logs of observations for the second set of samples.}
  \begin{center}
\begin{tabular}{lllllllll}
\hline \noalign{\vskip2pt} 
Target Name & Target ID$^a$ & R.A.$_{{\rm J2000}}$     &   Dec.$_{{\rm J2000}}$   & SpT$^b$   & T$_{\rm{eff}}$$^c$ & dist.$^c$ &  Obs. ID$^d$     &  Exp.$^e$  \\
  &          & [deg.]     & [deg.]                    &  & [K]   &  [pc]  &  &  [ks]     \\ \hline
Target 3$^f$  & \shortstack{\\3XMMJ000341.6-282347 \\(TYC 6418-1222-1)} & 0.924 & -28.397  & G8.5V & 5477 & 40 & 0602830101    &   7 \\ \hline
Target 4  & \shortstack{\\3XMMJ033201.9-522825 \\(TYC 8066-258-1)} & 53.009  & -52.474  & G3/5V & 5656 & 49 & 0400130101                          &    56     \\ \hline
Target 5  & \shortstack{\\3XMMJ051034.1-511803 \\(TYC 8084-1153-1)} & 77.642 & -51.301   & G6V & 5724 & 55 & 0729161001                          &     26     \\ \hline
Target 6  & \shortstack{\\3XMMJ085417.7-052603 \\(TYC 4873-1792-1)} & 133.573 & -5.434   & G2V & 5737 & 17 & 0404920201                          &     15     \\ \hline
Target 7  & \shortstack{\\3XMMJ124238.9+023436 \\(TYC 293-21-1)} & 190.662 & 2.576   & G8V & 5393 & 31 & 0111190701                          &     53     \\ \hline
Target 8  & \shortstack{\\3XMMJ213617.3-542551 \\(TYC 8811-516-1)} & 324.073 & -54.431   & G0V & 5769 & 33 & 0200230201                          &     6     \\ \hline
Target 9  & \shortstack{\\3XMMJ234256.5+001422 \\(TYC 586-1018-1)} & 355.736 & 0.240   & G8V & 5381 & 46 & 0211280101                          &     20     \\ \hline
Target 10  & \shortstack{\\3XMMJ235804.5-000741 \\(TYC 5253-655-1)} & 359.519 & -0.128   & G5V & 5859 & 66 & 0303110801                          &     7     \\ \hline
\end{tabular}
\label{tab:obssummary2}
\begin{flushleft} 
\footnotesize{
\hspace{0.5cm}$^a$ Target name in the X-ray and optical catalogs tabulated in \citet{2016A&A...590A...1R} and \citet{2003AJ....125..359W}. \\
\hspace{0.5cm}$^b$ Spectral type based on the SIMBAD astronomical database. \\
\hspace{0.5cm}$^c$ Effective temperature and distance are shown in \citet{2016A&A...595A...1G,2018AJ....156...58B}, respectively.\\
\hspace{0.5cm}$^d$ The XMM-Newton observation identification. \\
\hspace{0.5cm}$^e$ Net exposure time of the PN detector after removal of periods of high background flaring.\\
\hspace{0.5cm}$^f$ The star and an M dwarf constitute a common proper-motion pair, but they are separated by $65$ arcseconds (1700--2600 AU) \citep{2007AJ....133..439C}. This physical distance is so large that our spectral analysis is not affected by the companion star.}
\end{flushleft}
  \end{center}
\end{table*}

\subsection{Observations and data reduction}\label{sec:obs}
The basic information about the observations for our targets are summarized in Tables~\ref{tab:obssummary1} and \ref{tab:obssummary2}. We collected all the data available for our targets in the XMM-Newton science archive, and found twelve sets of observations. We analyzed all the twelve datasets and derived coronal quantities. Data reduction was conducted for the European Photon Imaging Camera (EPIC) data in the standard manner in SAS. The data taken from the PN instrument in the EPIC was analyzed to utilize a larger effective area than that of the EPIC-MOS instrument and to reduce the systematic error between the detectors. We analyzed the data taken from the EPIC-MOS2 instrument only for the observation of 0720253501 (Target~1), because the target is included only in the field of view of this detector. We removed data points during the observational periods of high instrumental-origin background flares only if data points exceed the 2$\sigma$ range in the count rate distribution and the removal of flares leads to a significant change in the spectra. We confirmed that the best fit parameters are consistent with each other within a statistical error.

\subsection{Analysis and results}\label{sec:analysis}
In order to investigate spectroscopic characteristics of our targets, we obtained
the X-ray spectra from individual observations and performed a spectral fitting to them. 
We also examined the time variability during individual observations to check that stellar flares did not dominate the stellar X-ray luminosity. As for Target~1, multi-epoch observations were made, which allows us to study the time variability on a timescale of years.

The procedure of the spectral fitting is as follows.
We created redistributed matrix and ancillary response files 
by using the SAS tasks, rmfgen and arfgen, respectively.
The spectra for the targets are extracted from the circles with a radius of 30 arcseconds centered on the targets, 
while the background spectra are extracted from the surrounding annulus with an inner and outer radii of 45 and 60 arcseconds, respectively, in order to take into account the spatial variation of the background. We subtracted the background spectra from the target spectra to obtain the intrinsic emissions from the targets.
The flux in the annulus is much smaller than the flux of the targets. Nevertheless, we optimized the size of the annulus to effectively remove contamination.
Following previous studies \citep[see, e.g., literature in][]{2009A&ARv..17..309G}, we used one-, two- or three-temperature, optically-thin thermal plasma models based on the collisional ionization equilibrium assumption, using the astrophysical plasma emission code, APEC \citep{2001ApJ...556L..91S}.
In our spectral analysis, the absorption and abundance parameters are set to be free. Solar photospheric abundances from \citet{1989GeCoA..53..197A} are used. We adopted the two- or three-temperature plasma models only if the models improves the fitting with a significance level of $\gtrsim$99\%.
Our two- and three-temperature models assume that the plasmas with different temperatures have the same abundance and absorption values.
To calculate the X-ray luminosity from the observed X-ray flux, we used the distance obtained by {\it Gaia} \citep{2018AJ....156...58B}. The uncertainty of the X-ray luminosity due to the error in distance is within 1 \%.

\begin{figure*}[h!]
\begin{tabular}{ccc}
\begin{minipage}{0.45\hsize}
\hspace{0.5cm}
(a) Target~1 (Obs. ID:0677181001)
\vspace{-0.3cm}
\begin{center}
    \includegraphics[width=0.85\linewidth]{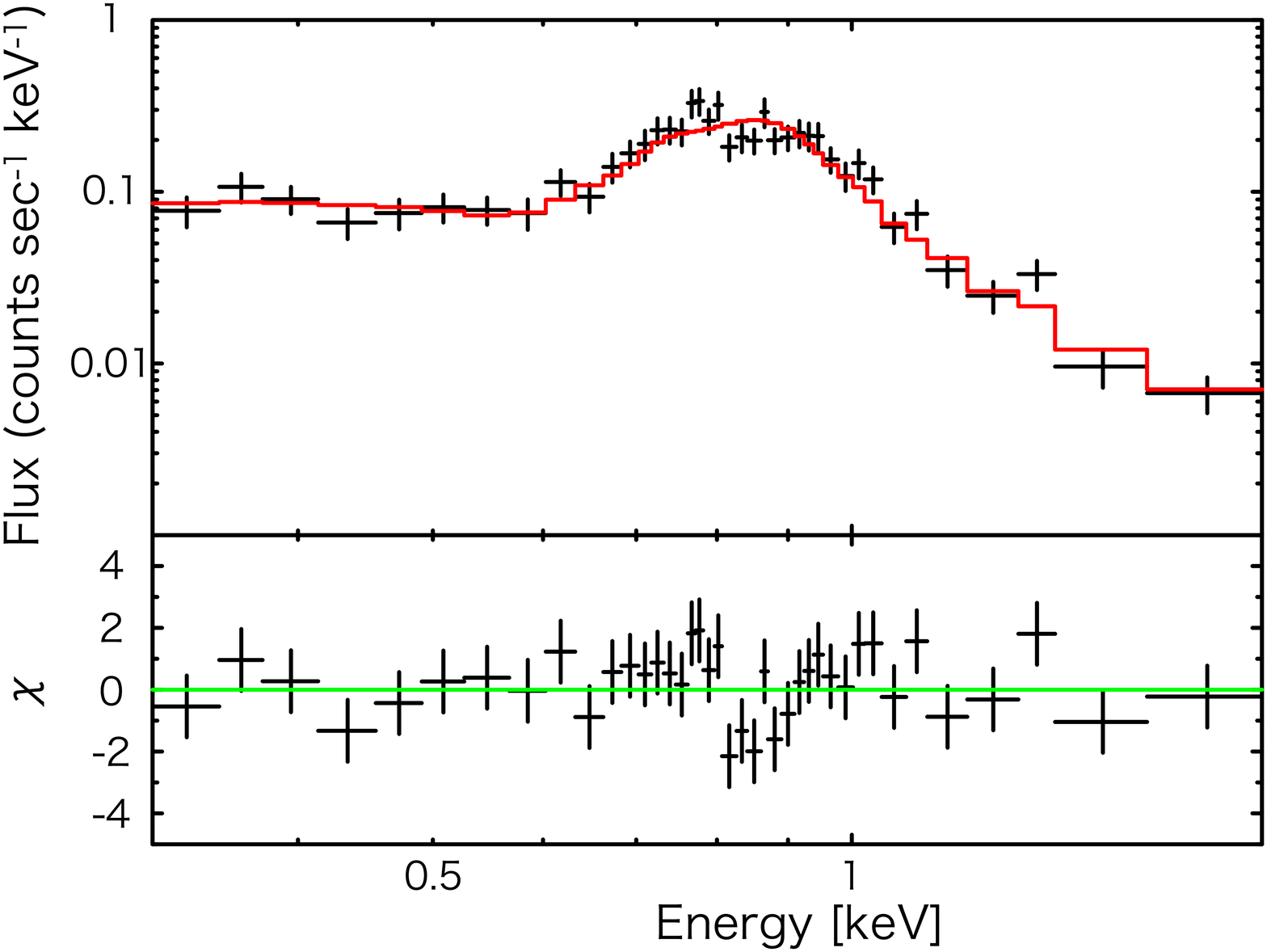}
\end{center}
\end{minipage}
\begin{minipage}{0.45\hsize}
\hspace{0.5cm}
(b) Target~1 (Obs. ID:0720251501) 
\begin{center}
\vspace{-0.3cm}
    \includegraphics[width=0.85\linewidth]{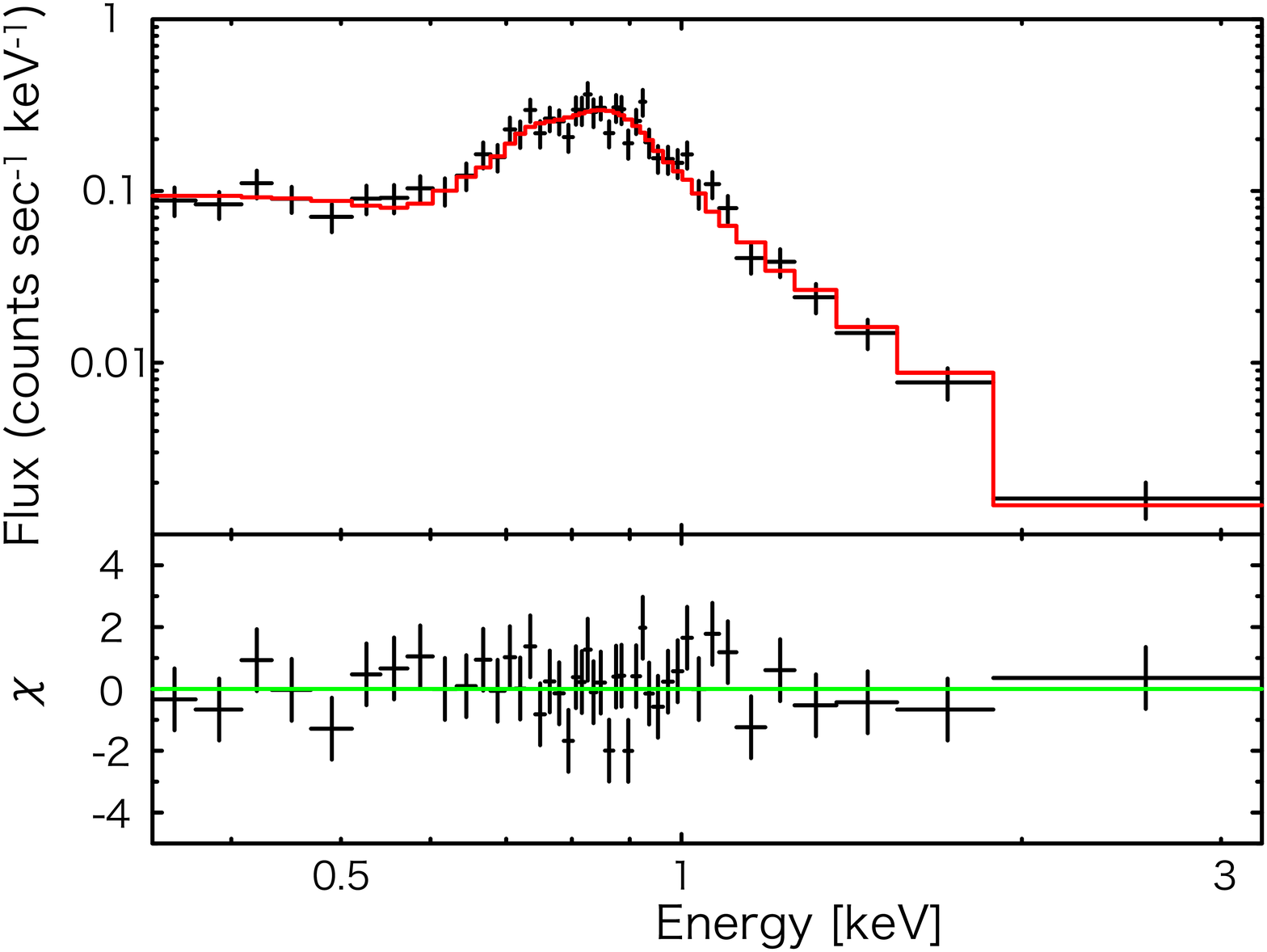}
\end{center}
\end{minipage}
\vspace{0.2cm}
\\
\begin{minipage}{0.45\hsize}
\hspace{0.5cm}
(c) Target~1 (Obs. ID:0720253501)
\begin{center}
\vspace{-0.3cm}
    \includegraphics[width=0.85\linewidth]{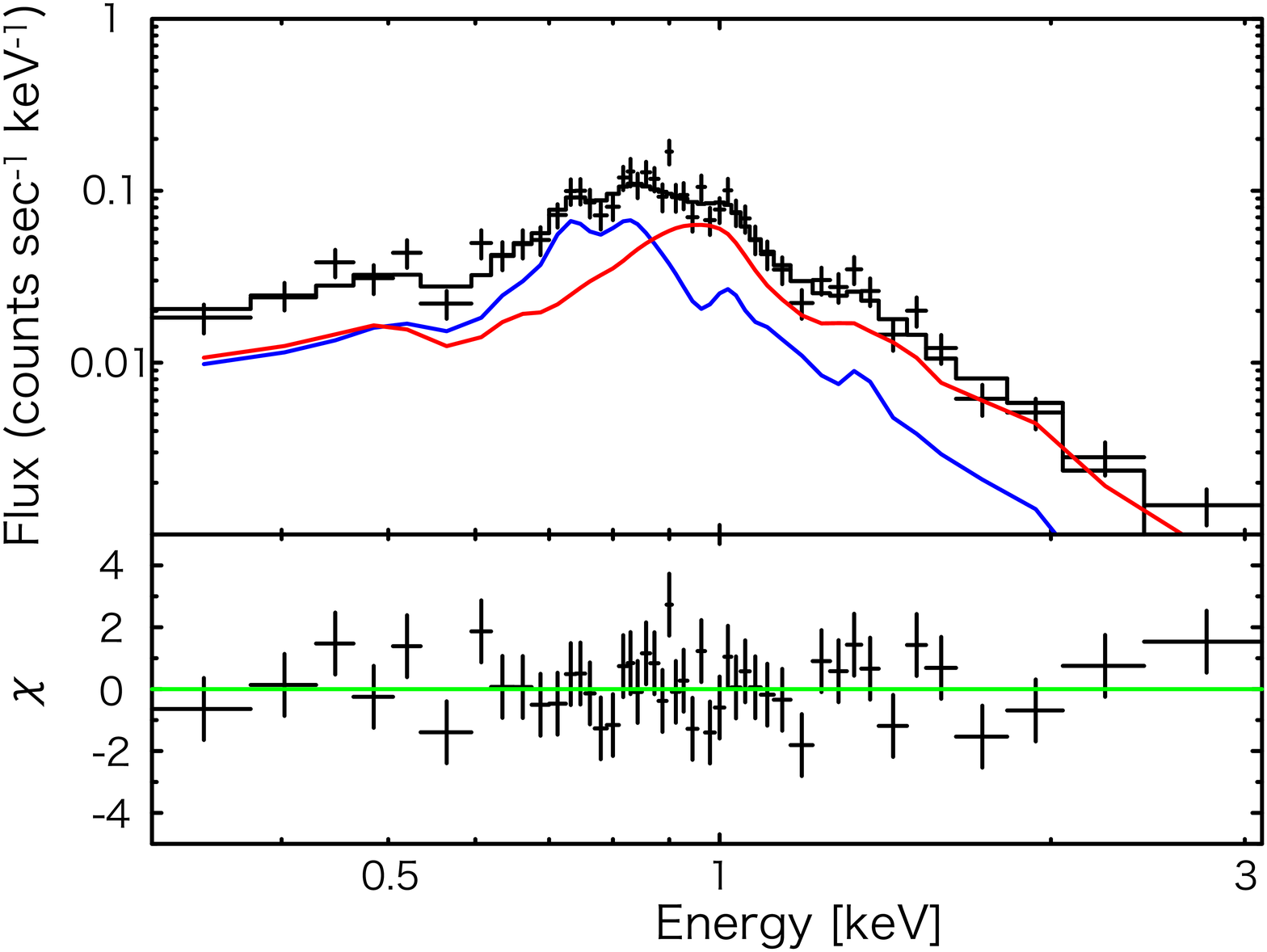}
\end{center}
\end{minipage}
\vspace{0.2cm}
\begin{minipage}{0.45\hsize}
\hspace{0.5cm}
(d) Target~2 
\begin{center}
\vspace{-0.3cm}
    \includegraphics[width=0.85\linewidth]{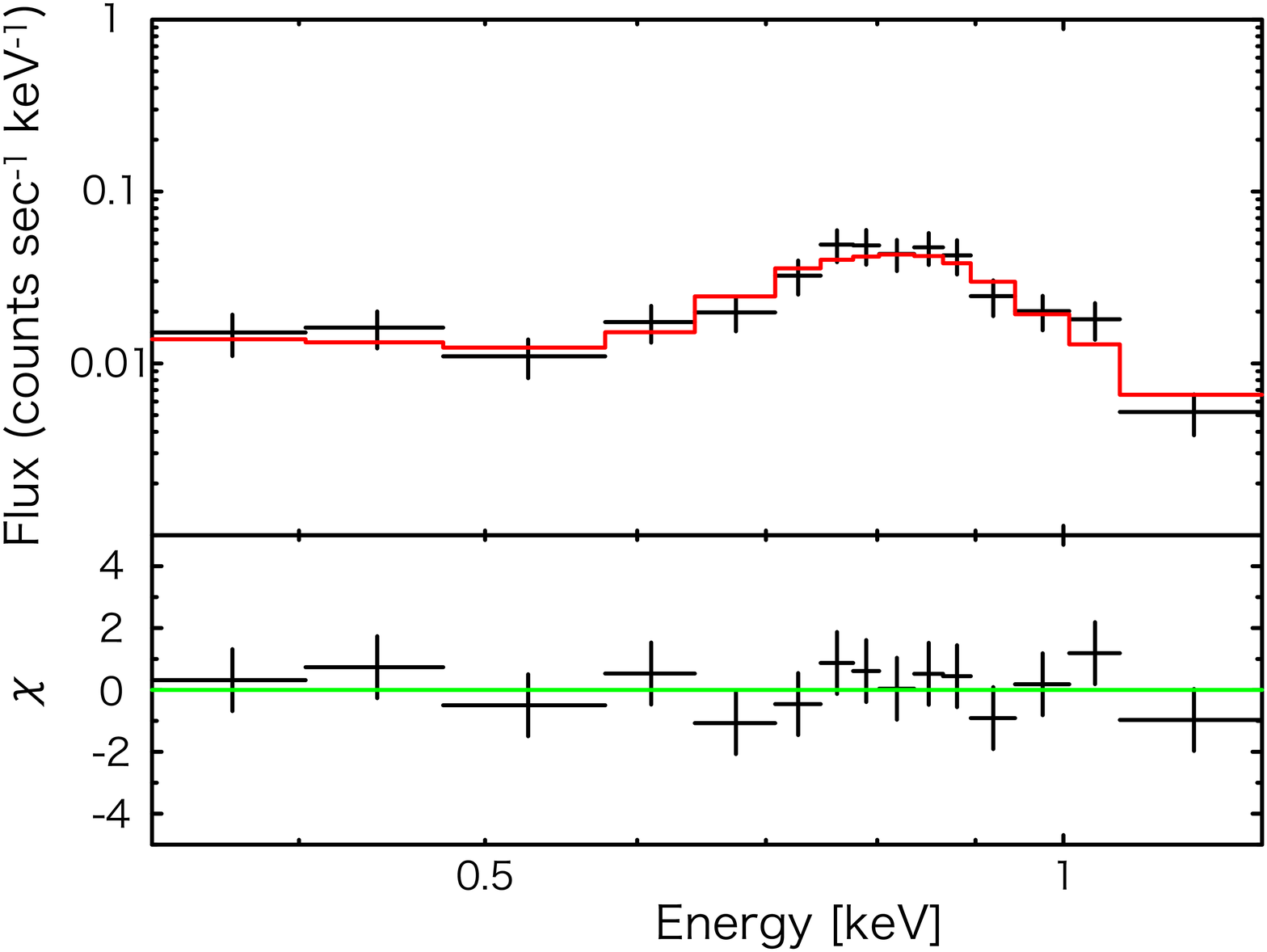}
\end{center}
\end{minipage}
\end{tabular}
  \caption{Spectra with the best fit models for Target~1 (a--c) and Target~2 (d). 
  For Target~1 (c), a two-temperature plasma model is applied. The cooler and hotter components are indicated by the blue and red lines, respectively. The spectra for all the other targets are fitted by a one-temperature plasma model (indicated by red lines).}
  \label{fig:spectra}
\end{figure*}

Figure~\ref{fig:spectra} and \ref{fig:spectra_add} show the observed energy spectra of the targets with spectral fits.
Results of the spectral fitting are summarized in Table~\ref{table:results} and \ref{table:results_add}.
We found that the one-temperature model is preferred except for one of the observations of Target~1 (3XMMJ041422.7-381900, Obs. ID: 0720253501).
The time variability is suggested in the catalogue only for Targets~4 and 6. To extract the spectra during the non-flaring periods, we examined both the light curves and the spectra for the two targets. We found no significant change in the spectra for Target~4. As for Target~6, we detected a hot plasma component only when the count rate was enhanced, which may suggest the occurrence of a stellar flare. Therefore, we do not include the data in this period for our spectral analysis.
For all the targets, the temperature for each plasma component is found to be $\sim$0.3--1 keV and the abundance is well constrained between $\sim$0.04 and 0.3 $Z_{\odot}$. The observed unabsorbed total X-ray luminosity ranges from $\sim 7\times10^{27}$ to $\sim 2 \times 10^{30}~{\rm erg~s^{-1}}$, while the total emission measure from $\sim  1\times10^{51}$ to $\sim 7\times 10^{52}~{\rm cm^{-3}}$.
We can see a positive correlation between the temperature and the emission measure as discussed later.
We confirmed that our spectral analysis for Targets~6 and 10 is consistent with the previous estimate of $L_{\it X}$ for those stars \citep{2014MNRAS.441.2361V,2009ApJS..181..444A}.

We also conducted additional spectral analysis to evaluate systematic errors. The systematic errors may arise from the difference of an abundance table, a plasma code, and contamination from a higher temperature plasma we cannot detect in our spectral analysis due to poor photon statistics above $\sim$1 keV. We use the table of \citet{asplund2009} and the plasma code \citet{mekal}, known as the mekal model, as alternative abundance table and plasma code, respectively.
We confirmed that all of the derived values are consistent with each other within a statistical error, except for the temperature with the mekal model. The temperature derived from the mekal model are systematically lower by 10--20\% than those of the obtained with the apec model as reported in previous studies (e.g., \citet{Gudel1997}.
To evaluate the effect of the contamination from a higher temperature plasma, we allowed plasmas with a temperature of $>$2~keV to be present when we search for the best fit models. We confirmed that all the parameters of interest remain unchanged within a statistical error regardless of the existence of the hot component. The results are shown in Tables~\ref{tab:obssummary1} and \ref{tab:obssummary2}.

The abundance for the X-ray bright stars are found to be significantly lower than the solar photospheric abundance. This is consistent with previous observations toward active stars. The depletion of elements with a smaller First Ionization Potential (FIP) is commonly seen in highly magnetically active stars \citep{brinkman2001,drake2001,gudel2002c}.

The stellar rotational periods are briefly discussed in Section~\ref{sec:rotation}. The period for Target~1 is estimated from the lightcurve taken by the {\it Transiting Exoplanet Survey Satellite} ({\it TESS}; \cite{ricker2014}), and the period for Target~6 is estimated in previous studies \citep[e.g.,][]{2018A&A...616A.108B,2014MNRAS.441.2361V}. We could not derive the reliable periods for the other sources.

\mitsuishi{
\begin{figure*}[h!]
\begin{tabular}{ccc}
\begin{minipage}{0.45\hsize}
\hspace{0.5cm}
(a) Target~3
\vspace{-0.3cm}
\begin{center}
    \includegraphics[width=0.85\linewidth]{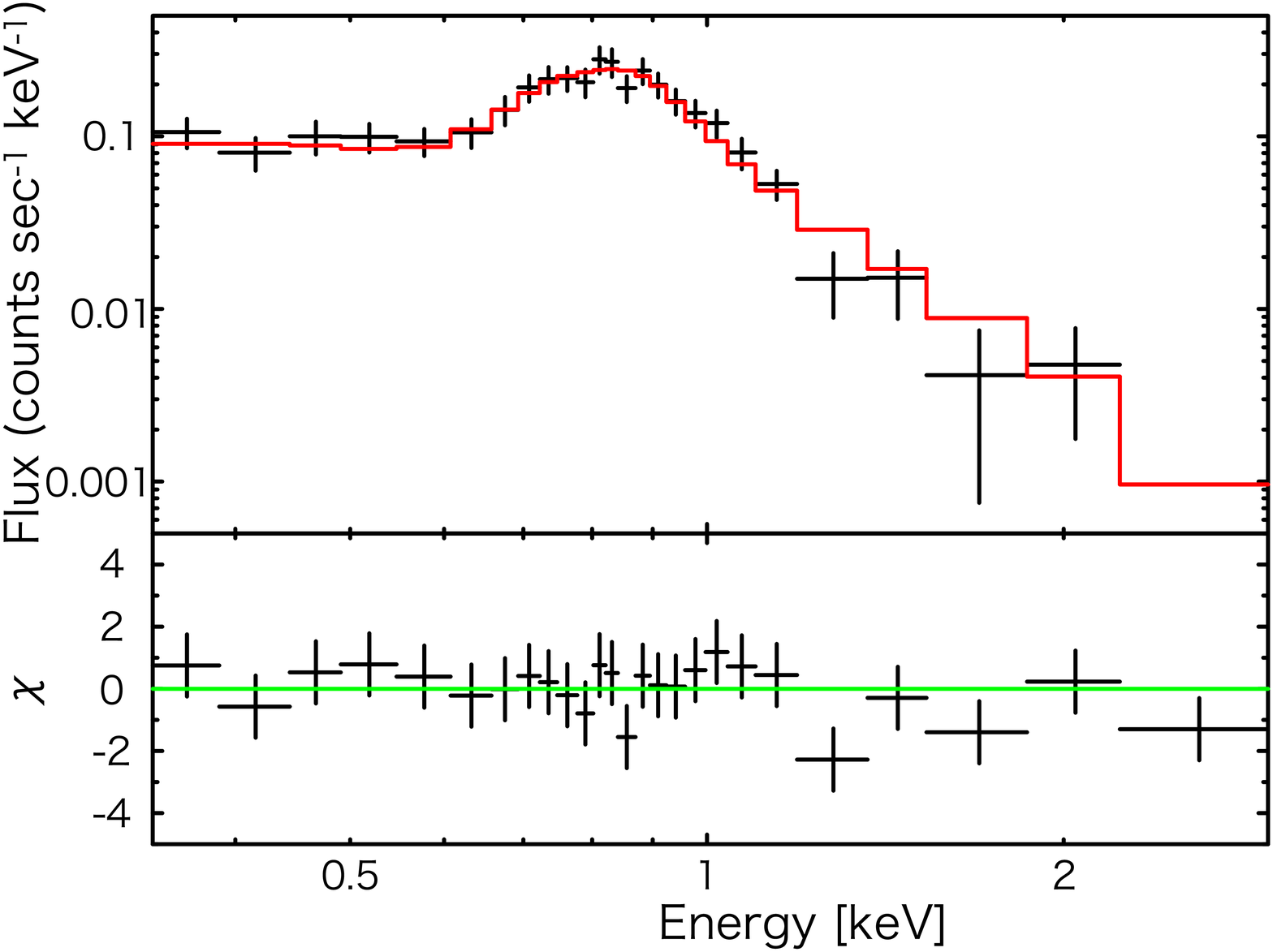}
\end{center}
\end{minipage}
\begin{minipage}{0.45\hsize}
\hspace{0.5cm}
(b) Target~4  
\begin{center}
\vspace{-0.3cm}
    \includegraphics[width=0.85\linewidth]{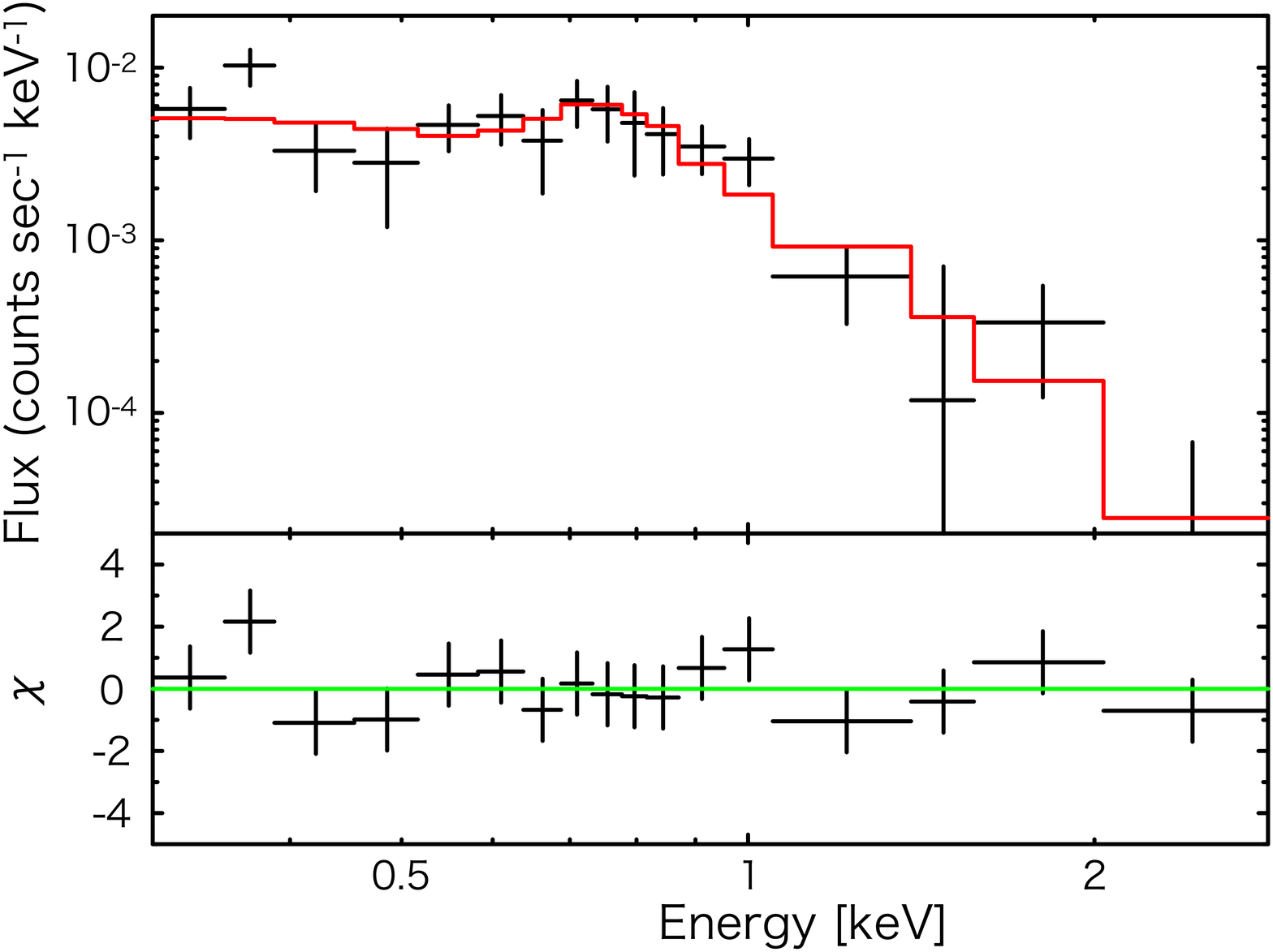}
\end{center}
\end{minipage}
\vspace{0.2cm}
\\
\begin{minipage}{0.45\hsize}
\hspace{0.5cm}
(c) Target~5
\begin{center}
\vspace{-0.3cm}
    \includegraphics[width=0.85\linewidth]{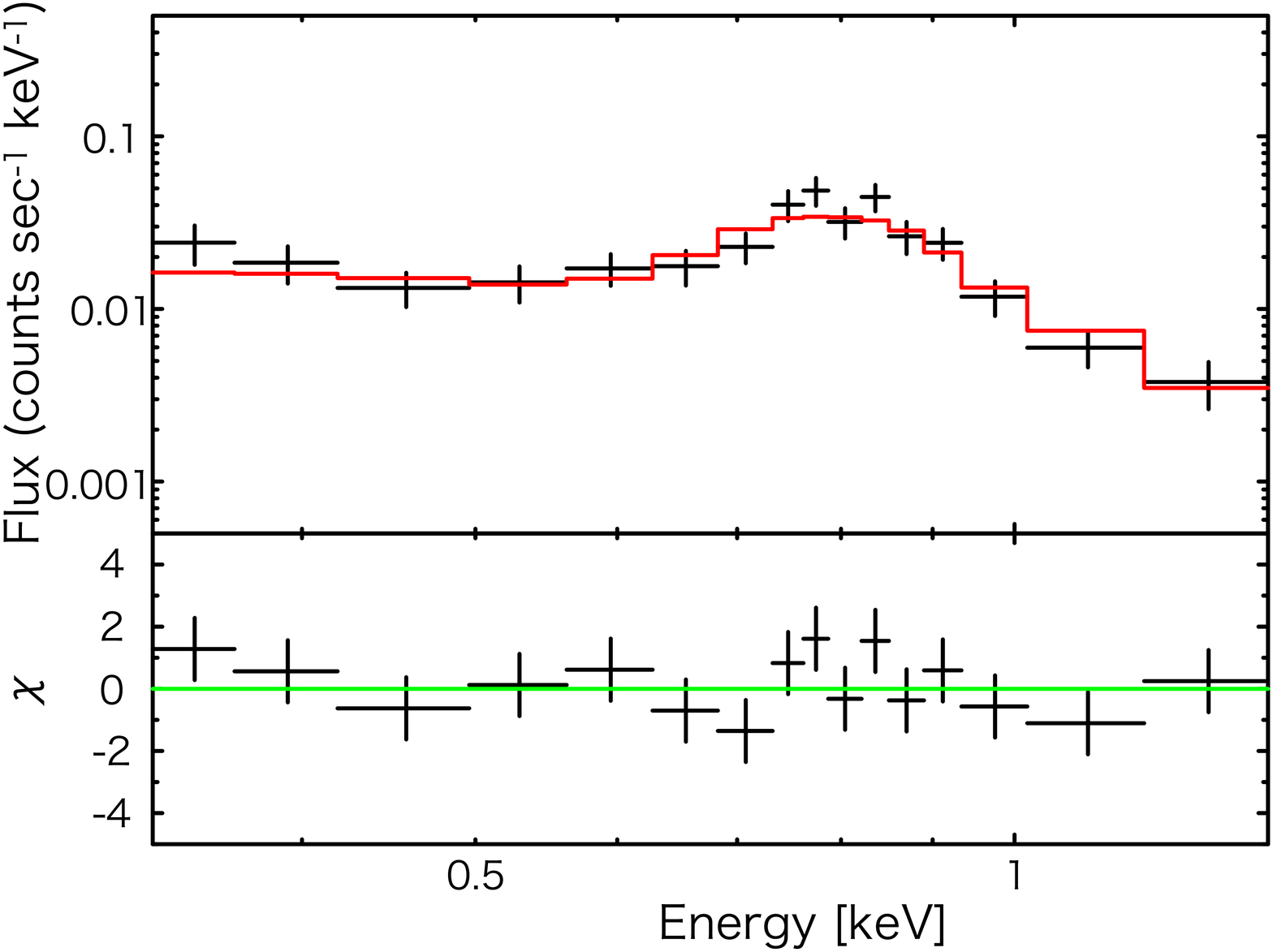}
\end{center}
\end{minipage}
\vspace{0.2cm}
\begin{minipage}{0.45\hsize}
\hspace{0.5cm}
(d) Target~6
\begin{center}
\vspace{-0.3cm}
    \includegraphics[width=0.85\linewidth]{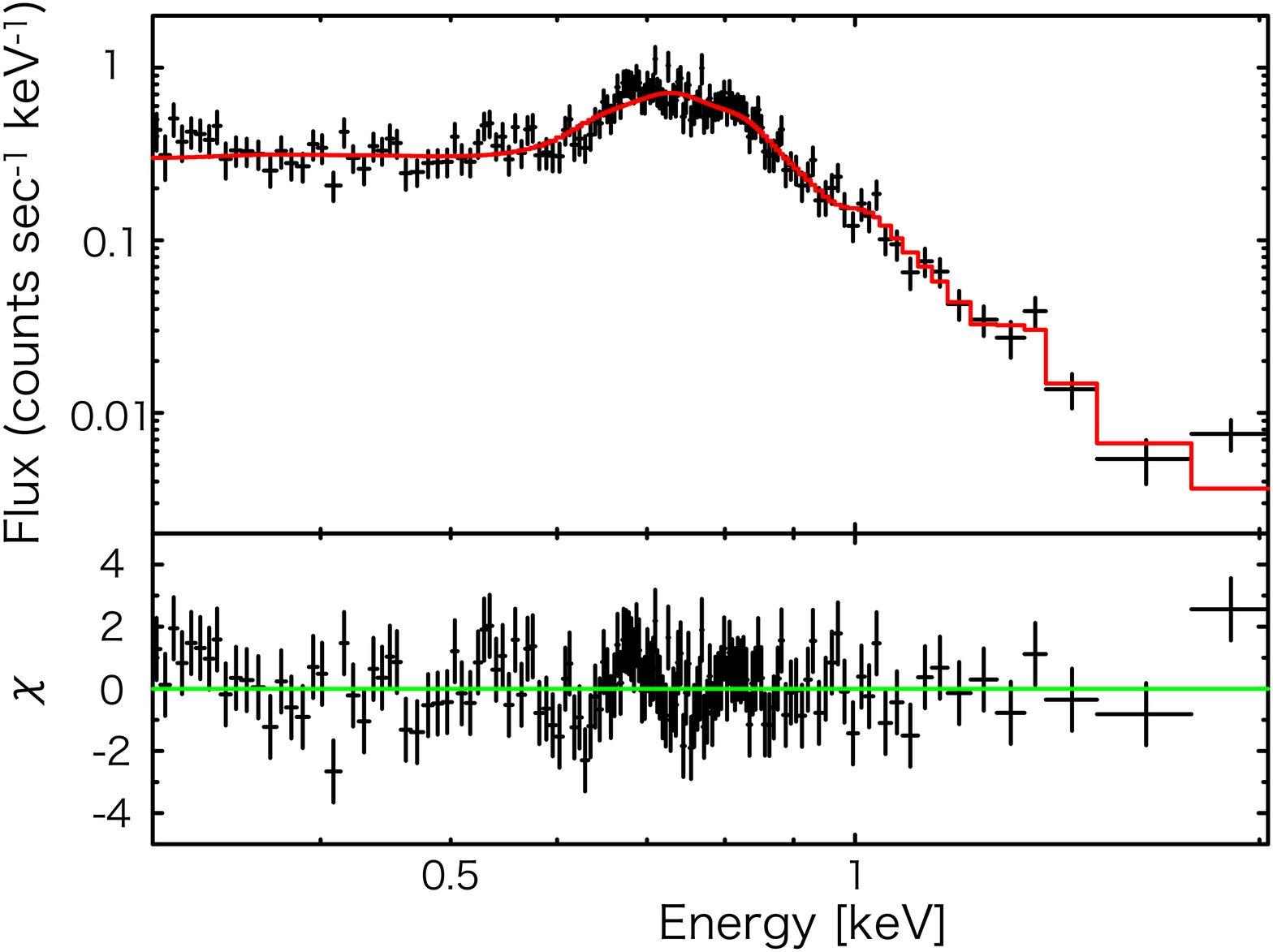}
\end{center}
\end{minipage}
\vspace{0.2cm}
\\
\begin{minipage}{0.45\hsize}
\hspace{0.5cm}
(e) Target~7 
\begin{center}
\vspace{-0.3cm}
    \includegraphics[width=0.85\linewidth]{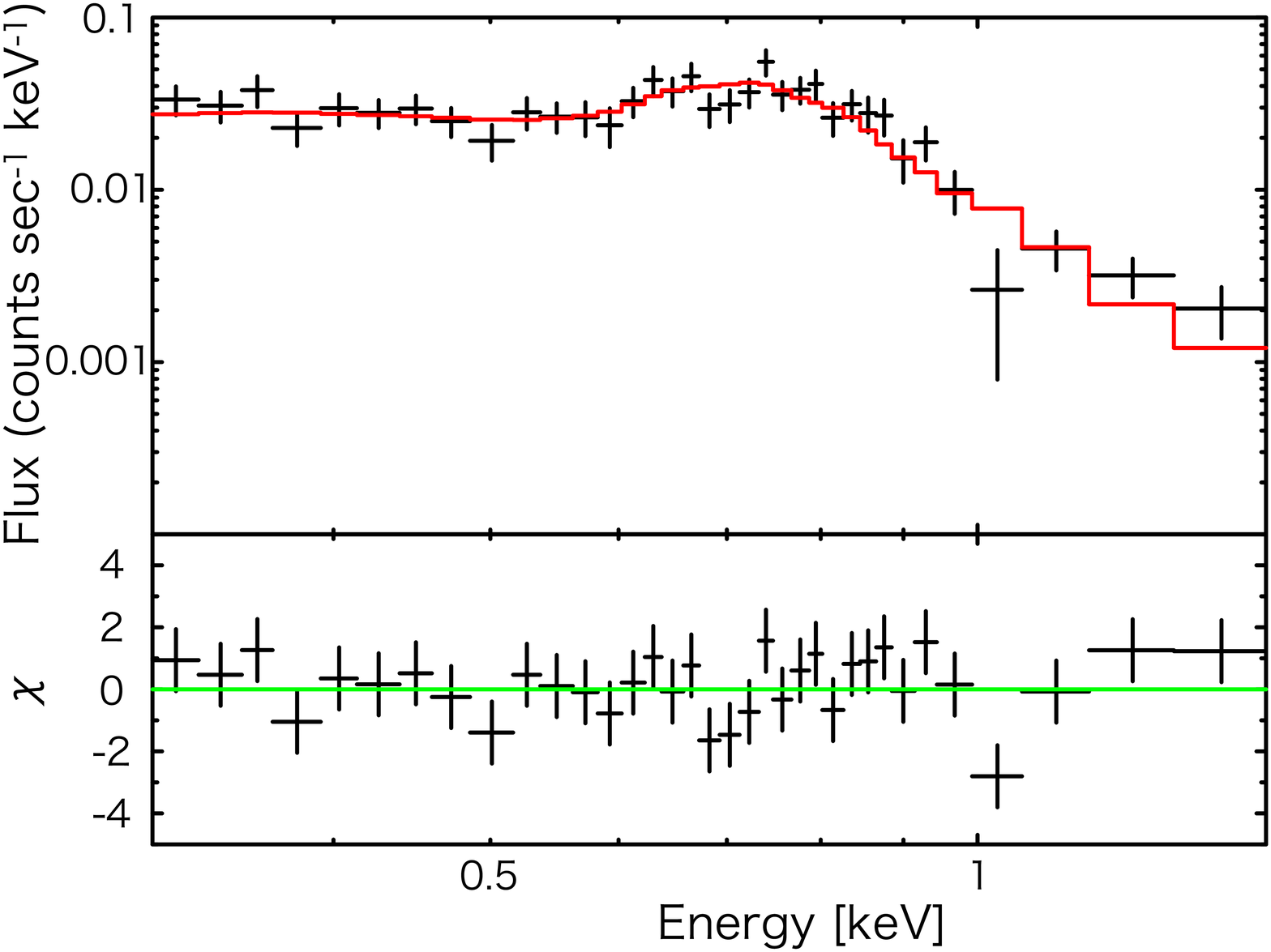}
\end{center}
\end{minipage}
\begin{minipage}{0.45\hsize}
\hspace{0.5cm}
(f) Target~8
\begin{center}
\vspace{-0.3cm}
    \includegraphics[width=0.85\linewidth]{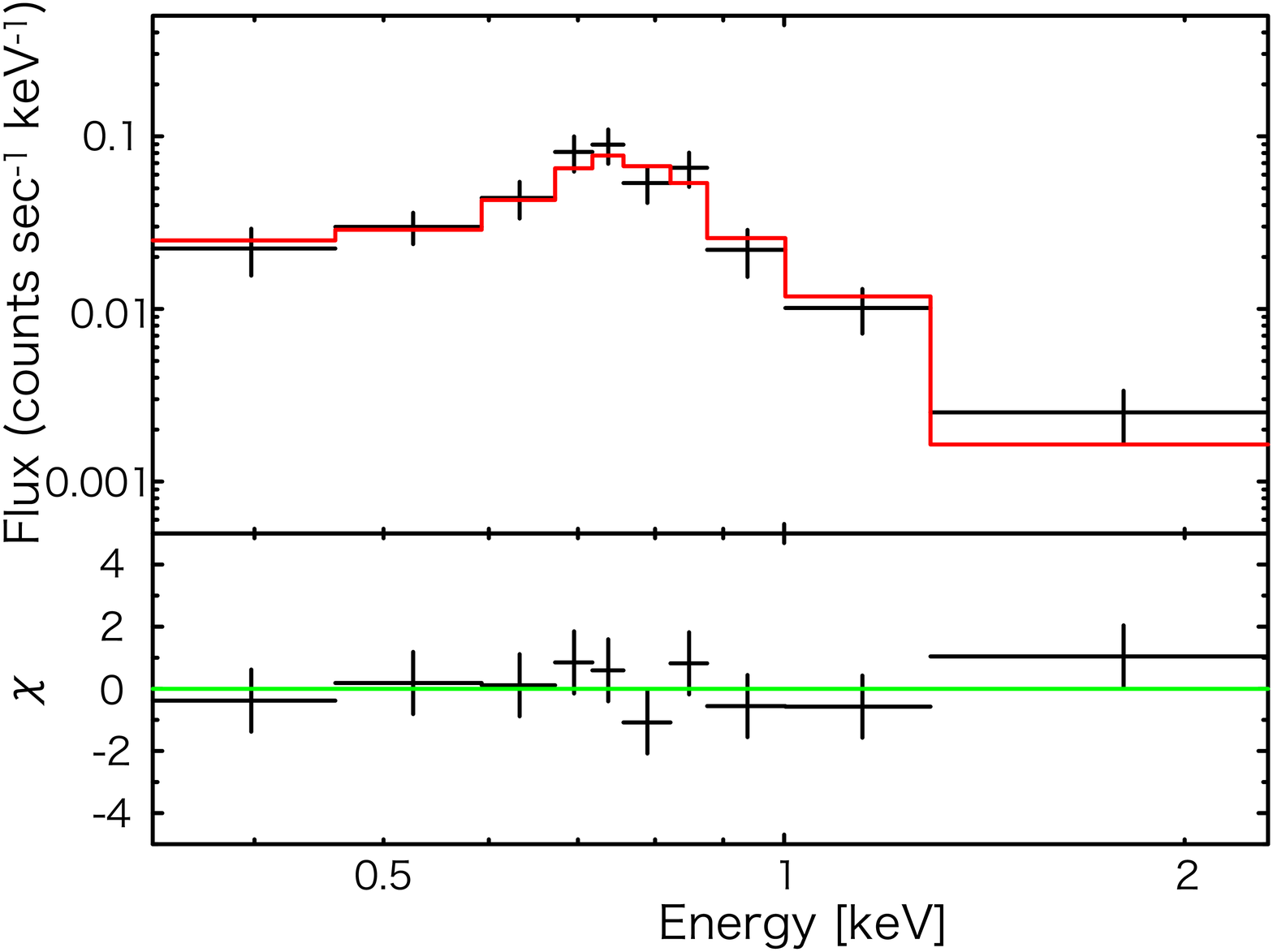}
\end{center}
\end{minipage}
\vspace{0.2cm}
\\
\begin{minipage}{0.45\hsize}
\hspace{0.5cm}
\vspace{0.2cm}
(g) Target~9
\begin{center}
\vspace{-0.3cm}
    \includegraphics[width=0.85\linewidth]{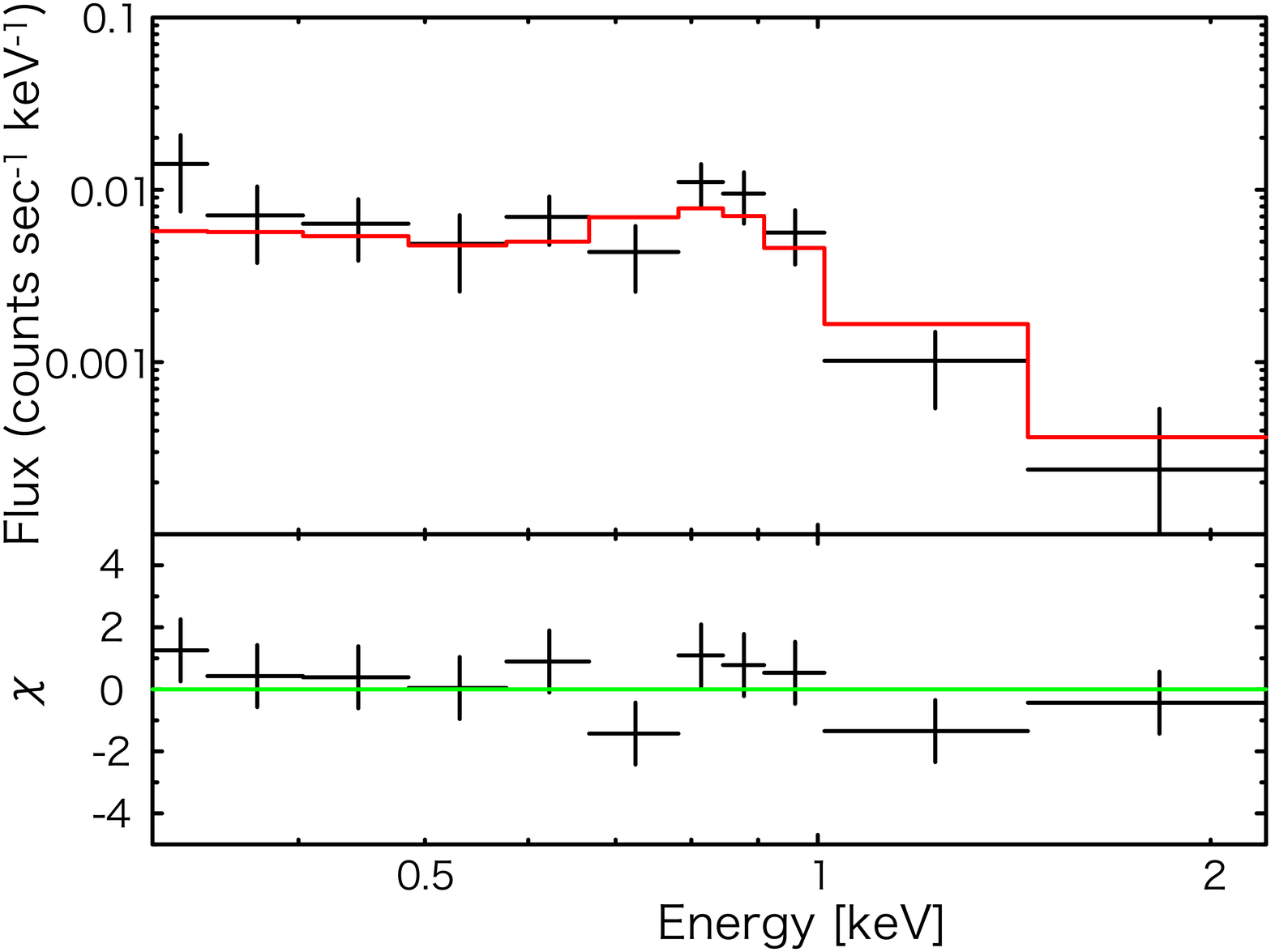}
\end{center}
\end{minipage}
\begin{minipage}{0.45\hsize}
\hspace{0.5cm}
(h) Target~10
\begin{center}
\vspace{-0.3cm}
    \includegraphics[width=0.85\linewidth]{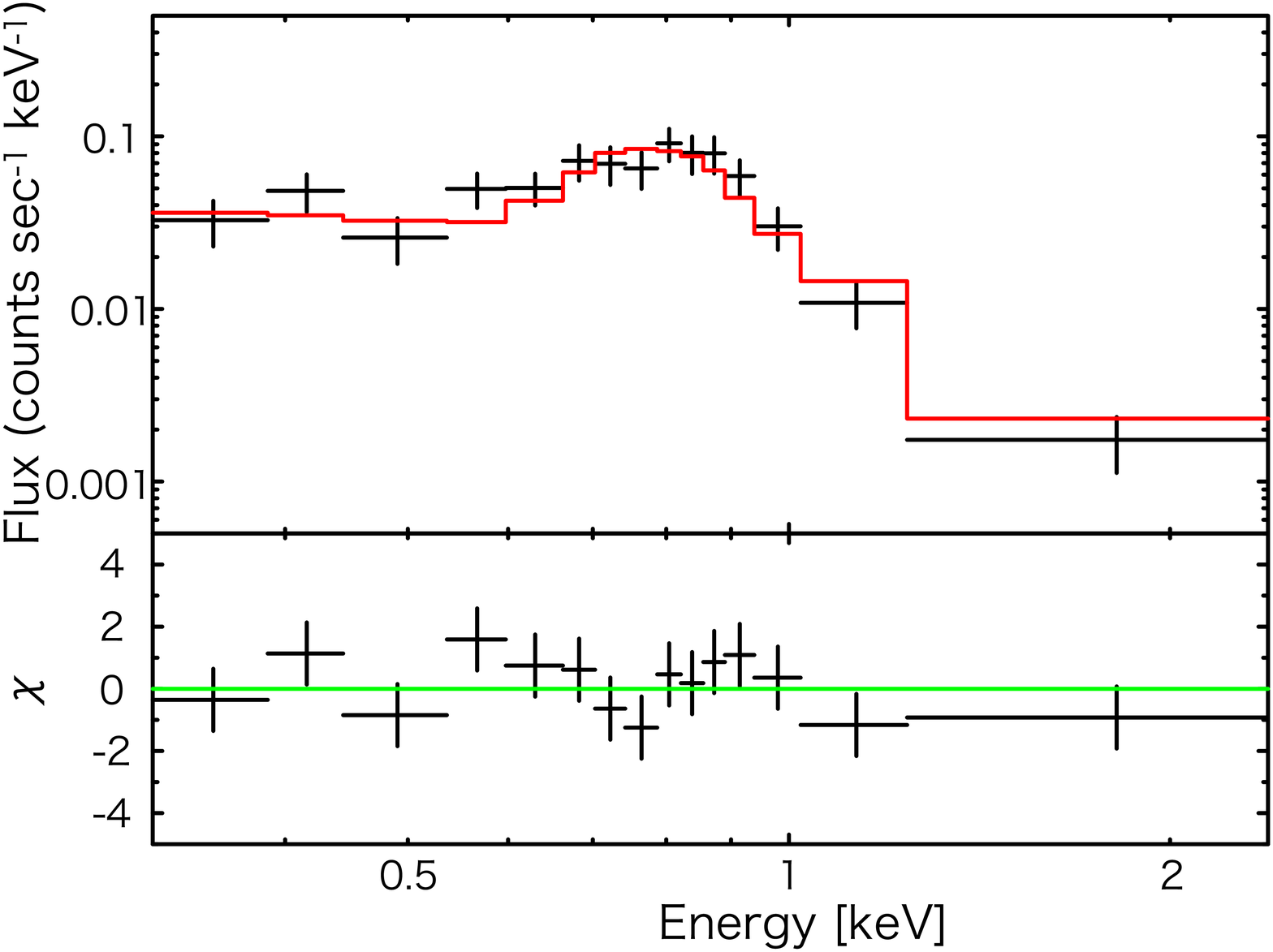}
\end{center}
\end{minipage}
\end{tabular}
  \caption{The same as Figure \ref{fig:spectra} but for the second set of samples.}
  \label{fig:spectra_add}
\end{figure*}
}

\mitsuishi{
\begin{table*}[h!]
\scriptsize{
  \caption{The best fit parameters of the model fitting for the X-ray bright stars.}
\label{table:results}
  \begin{center}
    \begin{tabular}{lllllllllll}
\hline\hline
Target Name      & Date   & $kT_\mathrm{low}$ & $EM_\mathrm{low}$ & $L_{\rm X, low}$ & $kT_\mathrm{high}$     & $EM_\mathrm{high}$ & $L_{\rm X, high}$     & $Z$ & $\chi^2/$d.o.f & $P_{\rm rot}$\\ 
      & yyyy/mm       &  [keV]          & [10$^{51}$cm$^{-3}$]          & [10$^{29}$~erg~s$^{-1}$]          & [keV]             &  [10$^{51}$cm$^{-3}$]         & [10$^{29}$~erg~s$^{-1}$]  &  [$Z_{\odot}$]              &   &  [day]\\ \hline \hline 
\multirow{9}*{Target~1}
& \multirow{3}*{2011/07$^a$} & & & & 0.76$\pm$0.04 & 52$^{+19}_{-8}$ & 7.9$^{+1.3}_{-0.5}$ & 0.27$\pm$0.07 & 45/36  & \multirow{9}*{1.8}
\\
& & & & & 0.64$\pm$0.03 & 61$^{+29}_{-21}$ & 8.5$^{+1.9}_{-1.2}$ & 0.31$^{+0.15}_{-0.10}$ & 46/36 &
\\
& & & & & 0.76$^{+0.04}_{-0.03}$ & 50$^{+21}_{-16}$ & 7.8$^{+1.4}_{-2.1}$ & 0.28$^{+0.13}_{-0.15}$ & 45/35  &
\\ \cline{2-10}
& \multirow{3}*{2013/07$^a$}  &   & &    & 0.75$^{+0.03}_{-0.04}$  & 58$^{+18}_{-10}$ & 8.8$^{+1.3}_{-0.4}$ & 0.27$^{+0.08}_{-0.07}$    & 38/40  &  
\\
& & & & & 0.62$\pm$0.03 & 73$^{+30}_{-24}$ & 9.9$^{+1.9}_{-1.7}$ & 0.30$^{+0.13}_{-0.08}$ & 33/40 &
\\
& & & & & 0.75$^{+0.03}_{-0.04}$ & 53$^{+23}_{-9}$ & 8.4$^{+1.7}_{-1.0}$ & 0.29$^{+0.12}_{-0.07}$ & 38/39 &
\\ \cline{2-10}
& \multirow{3}*{2014/02$^a$}  & 0.54$^{+0.14}_{-0.16}$  & 67$^{+45}_{-21}$ & 6.7$^{+4.5}_{-2.1}$   & 0.99$^{+0.17}_{-0.12}$  & 72$^{+40}_{-29}$ & 8.9$^{+3.9}_{-3.6}$ & 0.19$^{+0.09}_{-0.05}$    & 49/41  &
\\ 
& & 0.56$^{+0.07}_{-0.10}$  & 111$^{+64}_{-43}$ & 12$^{+5}_{-3}$   & $>$0.80  & 50$^{+99}_{-18}$ & 5.6$^{+4.3}_{-4.4}$ & 0.20$^{+0.17}_{-0.07}$    & 45/41 &
\\
& & 0.52$^{+0.15}_{-0.16}$  & 57$^{+47}_{-32}$ & 6.9$^{+3.6}_{-3.0}$   & 0.97$^{+0.08}_{-0.12}$  & 64$^{+43}_{-32}$ & 8.4$^{+4.3}_{-3.5}$ & 0.21$^{+0.20}_{-0.06}$    & 48/40  &
\\ \hline
%
\multirow{3}*{Target~2} & \multirow{3}*{2008/01$^a$}  &  &  & & 0.70$^{+0.06}_{-0.07}$ & 6.5$^{+5.2}_{-2.6}$    & 1.0$^{+0.4}_{-0.1}$ & 0.28$^{+0.23}_{-0.12}$    & 7/11 & \multirow{3}*{N/A}
\\
&  &  &  & & 0.58$^{+0.06}_{-0.07}$ & 6.1$^{+9.5}_{-2.9}$    & 1.0$^{+0.6}_{-0.1}$ & 0.38$^{+0.43}_{-0.21}$    & 8/11 &
\\
&  &  &  & & 0.70$^{+0.06}_{-0.07}$ & 6.7$^{+5.1}_{-3.5}$    & 1.0$^{+0.4}_{-0.2}$ & 0.27$^{+0.30}_{-0.14}$    & 7/10 &
\\ \hline
\hline
    \end{tabular}
  \end{center}
\begin{flushleft}
\footnotesize{
$^a$ Best fit parameters obtained by using $apec$ (first) and $mekal$ (second) plasma codes for the best fit 1$kT$ or 2$kT$ models. 
The best fit parameters with a high temperature plasma (2 keV) are also shown in the third line. 
}
\end{flushleft}}
\end{table*}
}

\begin{table*}[h!]
\scriptsize{
  \caption{The best fit parameters of the model fitting for additional solar-type stars to compare to the X-ray bright subsample.}
\label{table:results_add}
  \begin{center}
    \begin{tabular}{llllllll}
\hline\hline
Target Name      & Date   & $kT_\mathrm{high}$     & $EM_\mathrm{high}$ & $L_{\rm X, high}$     & $Z$ & $\chi^2/$d.o.f & $P_{\rm rot}$\\
                 & yyyy/mm         & [keV]             &  [10$^{51}$cm$^{-3}$]         & [10$^{28}$~erg~s$^{-1}$]  &  [$Z_{\odot}$]                 &     & [day] \\ \hline \hline
\multirow{3}*{Target~3$^a$} & \multirow{3}*{2009/12$^b$}  & 0.73$\pm$0.05 & 4.0$\pm$0.8    & 5.0$^{+0.6}_{-0.2}$ & 0.18$^{+0.07}_{-0.05}$    & 18/22 & \multirow{3}*{N/A}
\\
&  & 0.61$\pm$0.04 & 4.0$^{+2.0}_{-0.9}$    & 4.8$^{+1.2}_{-0.4}$ & 0.24$^{+0.10}_{-0.08}$    & 13/22 &
\\
&  & 0.73$^{+0.04}_{-0.05}$ & 4.0$^{+0.8}_{-1.0}$    & 5.0$\pm$0.4 & 0.18$^{+0.08}_{-0.05}$    & 18/21 &
\\ \hline
\multirow{3}*{Target~4} & \multirow{3}*{2006/5$^a$}  & 0.52$^{+0.19}_{-0.17}$ & 1.1$^{+0.9}_{-0.3}$    & 0.7$^{+0.4}_{-0.1}$ & 0.04$^{+0.07}_{-0.03}$    & 13/13 &\multirow{3}*{-}
\\
&  & 0.44$^{+0.17}_{-0.12}$ & 1.2$^{+1.5}_{-0.4}$    & 0.7$^{+0.6}_{-0.1}$ & 0.06$^{+0.08}_{-0.04}$    & 12/13 &
\\
&  & 0.53$^{+0.18}_{-0.25}$ & 1.1$^{+1.0}_{-0.7}$    & 0.7$^{+0.4}_{-0.1}$ & 0.04$^{+0.33}_{-0.03}$    & 13/12 &
\\ \hline
\multirow{3}*{Target~5} & \multirow{3}*{2016/11$^a$}  & 0.64$^{+0.06}_{-0.07}$ & 3.6$^{+1.4}_{-1.0}$    & 3.8$^{+0.8}_{-0.4}$ & 0.14$^{+0.08}_{-0.05}$    & 13/12 &\multirow{3}*{-}
\\
&  & 0.54$\pm$0.06 & 3.8$^{+2.0}_{-1.1}$    & 3.7$^{+1.0}_{-0.4}$ & 0.17$^{+0.10}_{-0.06}$    & 17/12 &
\\
&  & 0.64$^{+0.06}_{-0.07}$ & 3.7$^{+0.9}_{-1.3}$    & 3.8$^{+0.8}_{-0.4}$ & 0.13$^{+0.08}_{-0.04}$    & 13/11&
\\ \hline
\multirow{3}*{Target~6} & \multirow{3}*{2006/11$^a$}  & 0.37$\pm$0.01 & 2.2$\pm$0.3    & 2.1$\pm$0.1 & 0.21$^{+0.05}_{-0.04}$    & 153/142 &\multirow{3}*{15$^{c}$, 21$^{d}$}
\\
&  & 0.35$\pm$0.01 & 2.5$\pm$0.3    & 2.1$\pm$0.1 & 0.20$\pm$0.03    & 181/142 &
\\
&  & 0.37$\pm$0.02 & 2.1$\pm$0.4    & 2.1$\pm$0.1 & 0.22$^{+0.08}_{-0.04}$    & 153/141 &
\\ \hline
\multirow{3}*{Target~7} & \multirow{3}*{2000/07$^a$}  & 0.34$^{+0.05}_{-0.03}$ & 1.4$^{+0.6}_{-0.3}$    & 0.9$^{+0.3}_{-0.1}$ & 0.10$^{+0.05}_{-0.04}$    & 35/31 &\multirow{3}*{N/A}
\\
&  & 0.33$^{+0.03}_{-0.02}$ & 1.6$^{+0.7}_{-0.3}$    & 0.9$^{+0.2}_{-0.1}$ & 0.09$^{+0.04}_{-0.03}$    & 41/31 &
\\
&  & 0.33$^{+0.05}_{-0.03}$ & 1.2$^{+0.6}_{-0.4}$    & 0.8$^{+0.2}_{-0.1}$ & 0.12$^{+0.09}_{-0.06}$    & 35/30 &
\\ \hline
\multirow{3}*{Target~8} & \multirow{3}*{2004/05$^a$}  & 0.43$^{+0.15}_{-0.13}$ & 2.5$^{+7.4}_{-1.9}$    & 2.1$^{+3.8}_{-1.0}$ & 0.12$^{+0.42}_{-0.07}$    & 5/6 &\multirow{3}*{-}
\\
&  & 0.39$^{+0.11}_{-0.10}$ & 2.0$^{+8.3}_{-1.9}$    & 1.7$^{+3.9}_{-0.8}$ & $>$0.06    & 6/6 &
\\
&  & 0.38$^{+0.16}_{-0.09}$ & 0.8$^{+6.7}_{-0.7}$    & 1.3$^{+3.4}_{-0.6}$ & $>$0.06    & 4/5 &
\\ \hline
\multirow{3}*{Target~9} & \multirow{3}*{2005/06$^a$}  & 0.73$^{+0.15}_{-0.18}$ & 2.7$^{+1.8}_{-1.3}$    & 2.3$^{+0.7}_{-0.6}$ & 0.07$^{+0.14}_{-0.06}$    & 9/7 &\multirow{3}*{N/A}
\\
&  & 0.62$^{+0.14}_{-0.17}$ & 3.0$^{+3.4}_{-1.5}$    & 2.3$^{+1.2}_{-0.6}$ & 0.08$^{+0.15}_{-0.06}$    & 8/7 &
\\
&  & 0.73$^{+0.15}_{-0.21}$ & 2.7$^{+1.5}_{-1.4}$    & 2.3$^{+0.7}_{-0.6}$ & 0.07$^{+0.14}_{-0.05}$    & 9/6 &
\\ \hline
\multirow{3}*{Target~10} & \multirow{3}*{2006/06$^a$}  & 0.59$^{+0.08}_{-0.13}$ & 4.6$^{+3.2}_{-1.5}$    & 4.9$^{+1.6}_{-0.6}$ & 0.15$^{+0.12}_{-0.06}$    & 12/11 & \multirow{3}*{N/A}
\\
&  & 0.50$^{+0.08}_{-0.13}$ & 4.7$^{+7.6}_{-1.8}$    & 4.8$^{+3.8}_{-0.7}$ & 0.20$^{+0.17}_{-0.10}$    & 10/11 &
\\
&  & 0.60$^{+0.07}_{-0.14}$ & 4.5$^{+3.3}_{-1.4}$    & 4.9$^{+1.6}_{-0.6}$ & 0.16$^{+0.11}_{-0.06}$    & 12/10 &
\\ \hline
\hline
    \end{tabular}
  \end{center}
\begin{flushleft}
\footnotesize{
$^a$ The star and an M dwarf constitute a common proper-motion pair, but they are separated by $65$ arcseconds (1700--2600 AU) \citep{2007AJ....133..439C}. This physical distance is so large that our spectral analysis is not affected by the companion star.\\
$^b$ Best fit parameters obtained by using $apec$ (first) and $mekal$ (second) plasma codes for the best fit 1$kT$ or 2$kT$ models. 
The best fit parameters with a high temperature plasma (2 keV) are also shown in the third line.\\
$^c$ \citet{2018A&A...616A.108B}.\\
$^d$ \citet{2014MNRAS.441.2361V}.
}
\end{flushleft}}
\end{table*}

\section{Rotational periods of stars}\label{sec:rotation}
We briefly investigated the rotational periods of stars considered in this study. The period for Target~1 are derived from {\it TESS} observations. We estimated the stellar rotational period of our targets from their light curves by assuming that the quasi-periodic modulation is due to the presence of starspots. We computed a Lomb-Scargle periodogram for the stars to derive their rotational periods. The derived period is approximately 1.8~days.


The information about the stellar rotation and age was obtained only for some of our targets. Target~6 shows $v\sin i=1.2~{\rm km~s^{-1}}$ \citep{2008MNRAS.388...80P}. The age is estimated to be 2--4~Gyr \citep{2017ApJ...845...79B,2008MNRAS.388...80P}. The rotational period is estimated to be 15 to 21~days \citep{2018A&A...616A.108B,2014MNRAS.441.2361V}. Target~7 shows $v\sin i=4.9~{\rm km~s^{-1}}$ \citep{2006PASP..118..706P}. Target~9 shows $v\sin i=2.9~{\rm km~s^{-1}}$ \citep{2016ApJS..225...32B}. The age is estimated to be 5.4~Gyr \citep{2016ApJS..225...32B}. 
Namely, the rotation period is estimated only for Targets~1 and 6.

\begin{figure}
\epsscale{0.5}
\plotone{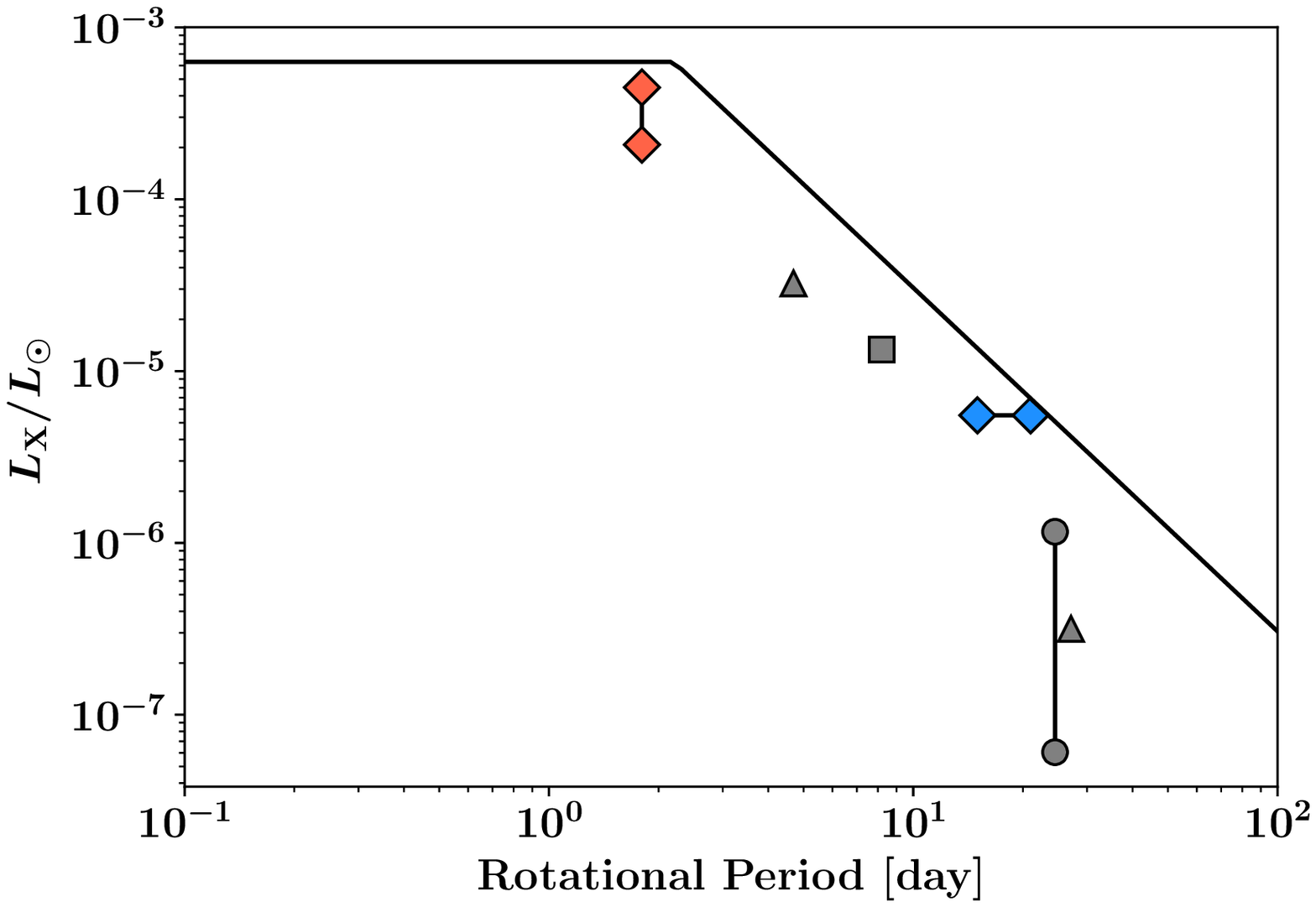}
\caption{$L_{\rm X}/L\odot$ vs rotational period diagram. Here only handful samples are plotted: Targets~1 (red) and 6 (blue), two stars from \citet{Gudel1997} (gray triangles), $\iota$ Horologii \citep{2019A&A...631A..45S} (gray square), and the Sun (gray circles, solar minimum and maximum). The gray triangle with the longer rotational period corresponds to $\beta$ Hyi. For the Target~6, we consider the uncertainty in the rotational period found in previous studies \citep[e.g.,][]{2018A&A...616A.108B,2014MNRAS.441.2361V}. The solid line shows the empirical relation from \citet{2003A&A...397..147P}.} \label{fig:LxLsun_period}
\end{figure}

Figure~\ref{fig:LxLsun_period} shows the $L_{\rm X}/L_\odot$-period diagram (since the stars in this study are all G-dwarf stars, we can assume $L_{\rm bol}\approx L_\odot$). Although only handful objects are plotted here, we can find that they basically follow the empirical relation from \citet{2003A&A...397..147P}. The X-ray brightest star, Target 1, seems to be in the X-ray saturated regime.

\section{Application of Derived Scaling Relation to Observations}\label{sec:theory-obs-comp}
We combine our samples with previously reported samples to cover a wide range of $L_{\rm X}$. We apply our theoretical scaling law (the combination of Equations~\ref{eq:EM-afterInteg} and \ref{eq:Tobs}) to the combined samples. The data for the solar minimum and maximum are taken from \citet{Peres2000} (filled gray circles). 
The data of a solar-analog, $\iota$ Horologii (HD~17051), is obtained from \citet{2019A&A...631A..45S} (gray square). They conducted a two-temperature fit toward multiple observations for a long-term period. Here, we adopt $T$ and EM for the higher temperature plasma component on 2014-02-05 only as a representative.
We also take the results of ROSAT observations toward two single stars from \citet{Gudel1997} ($\pi^1$ UMa and $\beta$ Hyi, namely HD 72905 and HD 2151, respectively. Gray triangles). 
We note that nine among 11 stars in \citet{Gudel1997} are binaries. They include spatially unresolved binaries in their analysis by assuming that both components of each binary are identical and contribute equally to $L_{\rm X}$. Here, we take a conservative approach; we include only the two single stars from their samples.

Before introducing our results, we note that $L_{\rm X}$ and $\rm EM_{tot}$ generally contain large intrinsic dispersion. The reasons are evident from X-ray and photospheric images of the Sun. Even in the solar maximum, the number of active regions is at most $\sim10$. In addition to this, active regions take various photospheric structures \citep[e.g.][]{kunzel1960} and their coronal activity is known to be sensitive to the photospheric structure \citep{sammis2000,2017ApJ...850...39T}. These factors introduce an intrinsic dispersion for coronal $L_{\rm X}$ and ${\rm EM_{tot}}$. The magnitude of the dispersion is hard to estimate, but if one looks at the well-known coronal activity--stellar rotation relationship \citep{2003A&A...397..147P}, one will find a dispersion of approximately an order of magnitude in $L_{\rm X}$ \citep{wright2011}. Therefore, in the following we will say that our scaling law agrees well with observations when the difference between the prediction and the observed value of $\rm EM_{tot}$ is within a factor of ten.


\begin{figure}
\epsscale{1.0}
\plotone{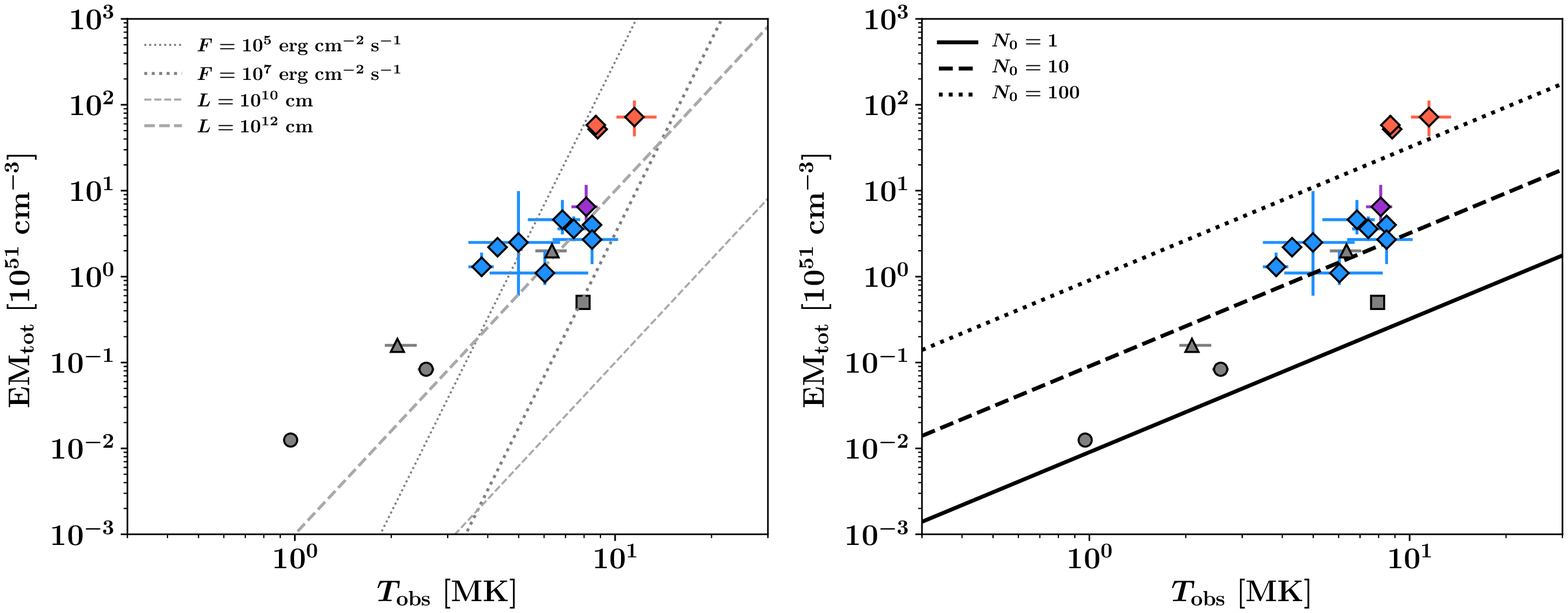}
\caption{${\rm EM_{tot}}-T_{\rm obs}$ relation. Target~1 and 2 are indicated by red and purple diamonds, respectively (See Table~\ref{tab:obssummary1}). Target~3 to 10 are shown by blue diamonds (See Table~\ref{tab:obssummary2}). Note that the best fit parameters in the first row (results from {\it apec}) in Tables~\ref{tab:obssummary1} and \ref{tab:obssummary2} are used. Two gray triangles denote data for G-dwarf stars in \citet{Gudel1997} (the results by the Raymond-Smith code \citep{1977ApJS...35..419R,1988ASIC..249....3R} are used), and the data with the lower EM is for $\beta$ Hyi. The gray square indicates the data for $\iota$ Horologii \citep{2019A&A...631A..45S}. Solar minimum and maximum are shown as gray filled circles \citep{Peres2000}. Left panel: Gray dashed and dotted lines show Equation~\ref{eq:EM-singleAR} with different parameters. Right panel: Data points with our scaling relation with different values of $N_0$ (see the text for the definition of $N_0$). Here we adopt $f=0.1$, $\bar{B}=100~{\rm G}$ and $F_{\rm h}=10^7~{\rm erg~cm^{-2}~s^{-1}}$.} \label{fig:EM-T}
\end{figure}

Figure~\ref{fig:EM-T} exhibits the ${\rm EM}_{\rm tot}$-$T_{\rm obs}$ relation. In the left panel, we also plot theoretical lines based on the single active region model by \citet{Shibata2002}: Equation~\ref{eq:l-T} (dashed lines, where we fixed $F_{\rm h}$ to $10^{7}~{\rm erg~cm^{-2}~s^{-1}}$) and Equation~\ref{eq:EM-singleAR-TF} (dotted lines, where we fixed $L$ to $10^{10}~{\rm cm}$), with different parameters. It is evident that the single active region model cannot fit most of the data points with reasonable parameter sets. For instance, one will find that the solar minimum and maximum are close to the line with $L=10^{12}$~cm. However, this spatial scale is larger than the solar radius, which is very unlikely. In addition, the solar data cannot be fitted by lines with a wide range of $F_{\rm h}$, say, $10^{5}~{\rm erg~cm^{-2}~s^{-1}}<F_{\rm h}<10^{7}~{\rm erg~cm^{-2}~s^{-1}}$.

The solid line in the right panel of Figure~\ref{fig:EM-T} shows our scaling law based on the present Sun. Here, we adopt $f=0.1$, $\bar{B}=100~{\rm G}$ and $F_{\rm h}=10^7~{\rm erg~cm^{-2}~s^{-1}}$. With these parameters, one will find that our scaling law is consistent with the data for the Sun, $\beta$ Hyi, and $\iota$ Horologii. We note that the rotational periods of these stars are $>8$~days. On the other hand, stars with much larger X-ray luminosity are located well above this line.

Previous observations of stellar spots suggest that rapidly-rotating stars show more starspots than slowly-rotating stars \citep{maehara2017}. Considering this suggestion, we plot our theoretical lines with $N_0=10$ (dashed) and $N_0=100$ (dotted) to quantitatively see the difference from the Sun-based ($N_0=1$) scaling law.
We find that our scaling law with $10\lesssim N_0 \lesssim 100$ may agree with the observations of X-ray bright stars.

\section{Summary and Discussion}\label{sec:summary-discussion}
The solar-stellar connection has been actively discussed recently \citep{2020Sci...368..518R,2020arXiv200613978Z}. Here, we studied the similarity and difference in coronal properties between the Sun and other G-dwarf stars, from both observational and theoretical points of view. 
We analyzed the X-ray archival data of single and wide binary G-dwarf stars. By combining the samples with data in literature, we investigated the coronal properties over a wide range of X-ray luminosity ($3\times 10^{26}$ to $2\times 10^{30}~{\rm erg~s^{-1}}$). For a quantitative comparison, we derived a scaling law of $\rm EM_{\rm tot}$--$T_{\rm obs}$ (the combination of Equations~\ref{eq:EM-afterInteg} and \ref{eq:Tobs}) for a star with multiple active regions, which links the surface magnetic fields and the coronal emissions. By applying the scaling law to the observations, we found that the scaling law is consistent with the data of the Sun, $\beta$ Hyi, and $\iota$ Horologii. However, the X-ray brighter stars show larger $EM_{\rm tot}$ than predicted from the Sun-based scaling law. Data points of such stars may be explained if the size distribution function of active regions is increased by 10--100 from the solar value.

We briefly discuss the limitations of our assumptions. We adopt the averaged field strength of 100~G for active regions in this paper. Although this averaged field strength is smaller than the typical field strength of the photospheric magnetic flux tubes, 1--2~kG (equipartition field, $B_{\rm eq}$), one will find that this value is consistent with observations of solar active regions by considering the effect of the filling factor $f_{\rm B}$, which is order of 0.1 \citep{1987A&A...180..241S}; namely, $\bar{B}=f_{\rm B}B_{\rm eq}\sim 100(f_{\rm B}/0.1)(B_{\rm eq}/1~{\rm kG})$~G. The assumption of $\bar{B}=100$~G for the averaged active region field strength will be valid if the averaged field strength over the whole stellar surface is well below this value. We can examine the condition from the $L_{\it X}$ vs total unsigned magnetic flux $\Phi_*$ relation by \citet{Pevtsov2003}, $L_{\rm X}\approx 10^{30}(\Phi_*/10^{26}~{\rm Mx})^{1.15}~{\rm erg~s^{-1}}$. When the averaged stellar field strength is 100~G, $\Phi_*\approx B A_\odot \approx 6\times 10^{24}$~Mx. The X-ray luminosity corresponding to this magnetic flux is $L_{\it X}\sim 10^{29}~{\rm erg~s^{-1}}$. Therefore, our assumption for the field strength of active regions will be valid for stars with $L_{\it X}\lesssim 10^{29}~{\rm erg~s^{-1}}$. More detailed considerations will be required for X-ray bright stars, although here we used the same field strength for such stars, Targets~1 and 2.

Another assumption we made is the same power-law index for the size distribution function as the solar value. Photospheric observations of solar-type stars suggest the size distribution of starspots is roughly located on the extension line of the distribution of sunspot groups \citep{ynotsu2019}. But the shape of the distribution function for very large starspots seems to be unconstrained very well yet, because of the low statistics. The distribution for large active regions will vary over a magnetic cycle as seen in \citet{parnell2009}.
In addition, the physical processes that determine the power-law index may be affected when the averaged stellar field strength is much larger than the solar value \citep[e.g.][]{2012ApJ...752..149I}. Further studies for magnetically very active stars will be therefore necessary. Detailed analyses about the starspot evolution will also be important to understand the distribution of surface magnetic fields \citep{namekata2019}. Continuous long-term observations are demanded to measure the shape of the distribution function. As for the Sun, investigations of the long-term solar activity with historical sunspot records will be useful \citep{2014SSRv..186...35C,2018PASJ...70...63H}.

Recently \citet{2020arXiv200613978Z} theoretically derived a power-law dependence of the X-ray emission on the total unsigned magnetic flux, using the RTV scaling. They obtained the dependence by averaging the physical quantities such as the magnetic field strength and the X-ray flux over the stellar surface, and neglected the spatial distribution of surface magnetic fields. Because of this treatment, there remains a degree of freedom in what we choose for the typical length for magnetic loops. They relate the power-law index with the physical background (e.g. averaged surface magnetic field strength and the heating model), but the coefficient of the power-law dependence (i.e., the absolute value) is not determined. In our multiple active region model, however, we consider the surface magnetic distribution and sum up the contribution from the active regions with different sizes. In this procedure, we care about the size of active regions, and therefore there is no ambiguity about the determination of the typical length of magnetic loops. As for our scaling law for the $\rm EM_{\rm tot}$--$T_{\rm obs}$ relation, we could determine not only the power-law index but also the coefficient, under our assumptions. Further consideration about e.g. the dependence of the heating flux on the stellar parameters will be required in our model.

The stellar rotational periods will be a key parameter for the $\rm EM_{\rm tot}$--$T_{\rm obs}$ relation, although the periods for most of our targets has not been determined in this study. Our Sun-based scaling law seems to be consistent with the data for stars with the period of $>$8~days, although Target~6 (period $\sim$15--21~days) is not fitted well. However, Target~1, the rapidly rotating star, is much brighter in X-ray than predicted by the Sun-based scaling law. Target~5 is also probably rapidly rotating (see Appendix), and Targets~7 and 9 may be so depending on their inclination angles (see Section~\ref{sec:rotation}). Although our sample number is quite limited, we infer from the above that this discrepancy may be explained if the rapidly rotating stars show more active regions with a given size than the Sun. Photospheric observations also suggested the possibility of the increase in starspots \citep{maehara2017,ynotsu2019}. We need more samples to clarify the relation among the stellar coronal activity, surface magnetic fields, and the rotation period.

\acknowledgments
We thank Drs. K. Shibata, T. K. Suzuki, Y. Notsu, M. Shoda, K. Masuda, and Mr. K. Namekata for fruitful comments on the solar-stellar connection. We are also grateful to Dr. K. Morihana. 
This work was supported in part by JSPS KAKENHI grant No. JP18K13579.
Based on observations obtained with XMM-Newton, an ESA science mission with instruments and contributions directly funded by ESA Member States and NASA. 
The ROSAT project was supported by the Ministerium f{\"u}r Bildung, Wissenschaft, Forschung und Technologie (BMBF/DARA) and by the Max-Planck-Gesellschaft. 
This research has made use of the SIMBAD database, operated at CDS, Strasbourg, France.
This work has made use of data from the European Space Agency (ESA) mission
{\it Gaia} (\url{https://www.cosmos.esa.int/gaia}), processed by the {\it Gaia}
Data Processing and Analysis Consortium (DPAC,
\url{https://www.cosmos.esa.int/web/gaia/dpac/consortium}). Funding for the DPAC
has been provided by national institutions, in particular the institutions
participating in the {\it Gaia} Multilateral Agreement.
This paper includes data collected by the TESS mission, which are publicly available from the Mikulski Archive for Space Telescopes (MAST).
This work made use of the IPython package \citep{PER-GRA:2007}, matplotlib, a Python library for publication quality graphics \citep{Hunter:2007}, and NumPy \citep{van2011numpy}.


\software{Numpy}

\begin{figure}
\epsscale{0.7}
\plotone{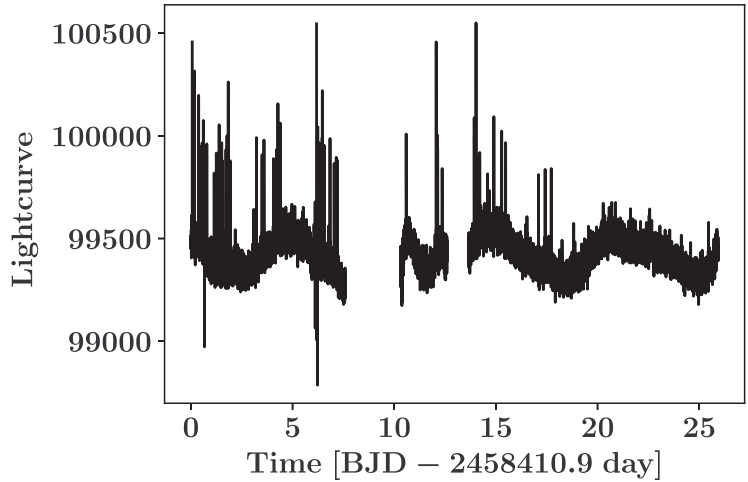}
\caption{{\it TESS} lightcurve for Target~5.} \label{fig:tess_target5}
\end{figure}

\appendix
{\it TESS} lightcurve for Target~5 is shown in Figure~\ref{fig:tess_target5} as a supplemental information of the stellar rotation period. The periodic change on a timescale of several days may be seen, although the observational period is insufficient to conclusively estimate the period. We also note that this star were producing many white-light flares.

\bibliography{StellarXray}

\end{document}